\documentclass[amssymb,amsmath,aps,showpacs,floatfix,nofootinbib,showpacs,12pt]{revtex4}
\usepackage{graphicx,epsfig}
\def\beq{\begin{equation}}
\def\eeq{\end{equation}}
\def\be{\begin{equation}}
\def\ee{\end{equation}}
\def\bea{\begin{eqnarray}}
\def\eea{\end{eqnarray}}
\def\nnb{\nonumber}

\newcommand{\gsim}{\lower.7ex\hbox{$\;\stackrel{\textstyle>}{\sim}\;$}}
\newcommand{\lsim}{\lower.7ex\hbox{$\;\stackrel{\textstyle<}{\sim}\;$}}

\begin{document}

%\newcommand{\Romannum4}{\uppercase\expandafter{\romannumeral4}} 
%\newcommand{\rmnum1}{\romannumeral1}
%\newcommand{\rmnum4}{\romannumeral4}

%\begin{center}
 %\vspace{0.2cm}
 \title{Effect of interaction with neutrons in matter
on flavor conversion of super-light sterile neutrino with active neutrino}
 \author{ Wei Liao, Yuchen Luo and Xiao-Hong Wu}
 \affiliation{
  Institute of Modern Physics, School of Sciences \\
 East China University of Science and Technology, \\
 130 Meilong Road, Shanghai 200237, P.R. China %
}

%\end{center}

\begin{abstract}
 %\vskip 0.2cm
A super-light sterile neutrino was proposed to explain the absence of
the expected upturn of the survival probability of low energy solar
boron neutrinos. This is because this super-light sterile neutrino can
oscillate efficiently with electron neutrino through a MSW resonance
happened in Sun. One may naturally expect that a similar resonance should
happen for neutrinos propagating in Earth matter. We study the flavor 
conversion of this super-light sterile neutrino with active neutrinos 
in Earth matter. We find that the scenario of the super-light sterile 
neutrino can easily pass through possible constraints from experiments 
which can test the Earth matter effect in oscillation of neutrinos. 
Interestinlgy, we find that this is because the naively expected resonant 
conversion disappears or is significantly suppressed due to the presence 
of a potential $V_n$ which arises from neutral current interaction of 
neutrino with neutrons in matter. In contrast, the neutron number density 
in the Sun is negligible and the effect of $V_n$ is effectively switched off. 
This enables the MSW resonance in Sun needed in oscillation of the super-light 
sterile neutrino with solar electron neutrinos. It's interesting to note that
it is the different situation in the Sun and in the Earth that makes $V_n$ 
effectively turned off and turned on respectively. This observation makes the 
scenario of the super-light sterile neutrino quite interesting.

\end{abstract}
 \maketitle

\newpage

 {\bf Introduction}

Among many candidates of sterile neutrino proposed in literature,
a super-light sterile neutrino appears to be very interesting~\cite{deHS0,deHS}.
With a mass squared difference with $\nu_1$ at around $\Delta m^2_{01} \approx
(0.5-2)\times 10^{-5}$ eV$^2$ and a mixing with electron neutrino 
around $\sin^2 2\theta_{01} \approx 0.001-0.005$~\cite{deHS,deHS1},
this sterile neutrino can help to explain the absence of the upturn
of the solar boron neutrino spectrum at energy $E_\nu \lsim 4$ MeV
\cite{Homestake,SK,SNOLETA,Borexino}
which is expected in the LMA MSW~\cite{w,ms} solution of the solar neutrino anomaly.
This is achieved with the help of a MSW resonant conversion of this
super-light sterile neutrino with solar electron neutrino when
neutrino travels from the interior of the Sun to the outside~\cite{deHS0,deHS}. 

One may naturally expect that there might exist a resonance of flavor
conversion between this sterile neutrino and electron neutrino 
when neutrinos propagate in Earth matter. If a resonance happen,
the effective mixing angle between this sterile neutrino and electron
neutrino can reach maximal and the oscillation phase would be around 
$V_e L \sin2\theta_{01}$ which can be of order one for a long enough
oscillation length $L$, e.g. for neutrinos crossing 
the core of the Earth. Hence, the probability of flavor conversion could be
large in resonance region. In particular, this would lead to a suppression of the 
total flux of active neutrinos when neutrinos pass through the core of the
Earth.

Since the mass squared difference $\Delta m^2_{01}$
is several to ten times smaller than the
solar mass squared difference $\Delta m^2_{21}$, the associated resonance may happen
at an energy much lower than the energy of the $1-2$ resonance which
is around $100-200$ MeV in Earth matter. 
In particular, this resonant conversion of sterile neutrino with
electron neutrino may happen for high energy solar neutrinos,
supernovae neutrinos and for low energy
atmospheric neutrinos. Hence, constraints on this super-light sterile neutrinos
may exist in low energy atmospheric neutrino data, supernovae neutrino data
and the test of the Earth matter effect in solar neutrino data.
In this article~\cite{others} we examine the oscillation of 
the super-light sterile neutrino
with active neutrinos in the Earth matter. Interestingly, we find that
the naively expected resonant conversion between this super-light sterile neutrino 
with active neutrino disappears or is significantly suppressed
in the presence of a sizeable
potential from the neutral current interaction with matter.
This is different from the oscillation happened in the Sun 
for which the neutron number density is low and
the effect from neutral current can be effectively neglected.
Consequently, we find that this super-light sterile neutrino passes the possible
constraints from experiments testing neutrino oscillation in Earth matter.

In the following of the present article we will first give a brief
review of the formalism and the convention of neutrino oscillation with a
super-light sterile neutrino. Then we will discuss the reduction
of the $4\nu$ formalism of neutrino oscillation in Earth matter
to a $3\nu$ formalism at energy $< 1$ GeV. In this $3\nu$ formalism
we will discuss in detail, with the help of a baseline dependent
average potential, the crossing of energy levels of neutrinos
and the oscillation of the super-light sterile neutrino
with active neutrinos. We will show how the naively expected
resonant conversion of this super-light sterile neutrino and active
neutrinos disappears in the presence of $V_n$, an effective potential
coming from neutral current interaction with neutrons in matter.
For the completeness of the discussion in this article,
we will then discuss the description of the oscillation of super-light
sterile neutrino using the baseline dependent average potential
and show that the discussion for the level-crossing using
this baseline dependent potential is indeed a valid and good
description despite the fact that Earth matter has complicated
density profile. Finally we conclude.

{\bf Super-light sterile neutrino and its oscillation in
$4\nu$ and $3\nu$ formalism}

In the presence of a super-light sterile neutrino, the Hamiltonian governing
the oscillation of $\nu=(\nu_s,\nu_e,\nu_\mu,\nu_\tau)^T$, the neutrino in 
flavor base, is
\bea
H=U H_0 U^\dagger+V, \label{H0}
\eea
where
\bea
&& H_0=
\frac{1}{2E} \textrm{diag}\{\Delta m^2_{01},0,\Delta m^2_{21},\Delta m^2_{31} \},
\label{H0a} \nnb \\
&& V=\textrm{diag} \{0, V_e+V_n,V_n,V_n \}. \label{H0b}
\eea
$V_e=\sqrt{2}G_F n_e$ and $V_n=-\frac{1}{\sqrt{2}} G_F n_n$ where
$n_e$ and $n_n$ are number densities of electron and neutron
in matter. $U$ is a $4\times 4$ unitary matrix describing the
mixing of neutrinos. Neglecting CP violating phases, it 
can be parameterized by
\bea
U=R(\theta_{23}) R(\theta_{13}) R(\theta_{12}) R(\theta_{02}) R(\theta_{01})
R(\theta_{03}) \label{H0c},
\eea
where $R(\theta_{ij})$ is a $4\times 4$ rotation matrix with a mixing angle 
$\theta_{ij}$ appearing at $i$ and $j$ entries, e.g.
\bea
R(\theta_{01})=\begin{pmatrix} \cos\theta_{01} & \sin\theta_{01} & 0 & 0
\cr -\sin\theta_{01} & \cos\theta_{01} & 0 & 0 \cr
 0 & 0 & 1 & 0 \cr 0 & 0 & 0 & 1 \cr \end{pmatrix}, \label{R01}
\eea
and
\bea
R(\theta_{02})=\begin{pmatrix} \cos\theta_{02} & 0& \sin\theta_{02} & 0 \cr
 0 & 1 & 0 & 0 \cr
 -\sin\theta_{02} & 0 & \cos\theta_{02} & 0  \cr
 0 & 0 & 0 & 1 \cr \end{pmatrix}. \label{R02}
\eea
$\theta_{12,13,23}$ are mixing angles governing the
flavor conversion of solar neutrinos, reactor neutrinos at short baseline
and atmospheric neutrinos separately and they have been measured
in solar, atmospheric, long baseline and reactor neutrino experiments
~\cite{RPP,theta13,theta13-2,theta13-3}.
If $\theta_{0i}(i=1,2,3)$ are all zero, the mixing matrix $U=
R(\theta_{23}) R(\theta_{13}) R(\theta_{12})$ and it reproduces the PMNS
mixing matrix without CP violating phase for three active neutrinos.
For anti-neutrinos the Hamiltonian is (\ref{H0}) with $V$ replaced by $ -V$.

As observed in \cite{deHS}, the presence of a super-light sterile
neutrino with $\Delta m^2_{01}\approx (0.5-2) \times 10^{-5}$ eV$^2$
and $\sin^22\theta_{01}\approx 0.001-0.005$ can lead to a further suppression
of the flux of the solar electron neutrinos at energy around $\lsim 4$ MeV.
Hence it provides an explanation of the absence of the upturn of the
solar neutrino flux at this energy range. For a small but non-zero $\theta_{02}$,
similar phenomena would occur for solar electron neutrinos~\cite{deHS}. 

For a small but non-zero $\theta_{03}$, solar electron neutrinos are
basically not affected by it but a 
resonant $\nu_s-\nu_\tau$(or ${\bar \nu}_s-{\bar \nu}_\tau$) oscillation 
of atmospheric neutrino is 
expected to happen for energy around $10-15$ GeV~\cite{deHS}.
So the scenario with a non-zero $\theta_{03}$ can be tested
in future atmospheric neutrino experiments and long baseline experiments.

In this article we study the effect of $U_{s1}$
and $U_{s2}$(basically $\sin\theta_{01}$ and $\sin\theta_{02}$) 
in oscillation of neutrinos in 
Earth matter. For energy as high as $\gsim 1$ GeV, oscillation of active neutrinos
with sterile neutrino due to effects of $U_{e1}$ and $U_{e2}$ would be
strongly suppressed since $\Delta m^2_{01}\approx (0.5-2)\times 10^{-5}$ eV$^2$
and $\Delta m^2_{21} \approx 7.5 \times 10^{-5}$ eV$^2$
and the oscillation lengths of $1-2$ and $0-1$
oscillation are all much longer than the diameter of the Earth for
such high energy.
Instead, we concentrate on oscillation of neutrinos
for energy $< 1$ GeV. In this energy range, the $2-3$ oscillation
is pretty fast and consequently we can reduce the $4\nu$ formalism 
to a $3\nu$ formalism in the study of the resonant oscillation between sterile neutrino
and active neutrino. We will see that this $3\nu$ formalism is more
convenient for later discussion.

The $3\nu$ formalism can be achieved from (\ref{H0}) as follows.
We can first rotate the Hamiltonian by $R(\theta_{23})$ and $R(\theta_{13})$.
Introducing $\nu'=(\nu_s,\nu'_e,\nu'_\mu,\nu'_\tau)^T$, which
is obtained by
\bea
\nu'=R(\theta_{23}) R(\theta_{13})\nu, \label{nuprime}
\eea 
we find that the Hamiltonian for $\nu'$ is
\bea
H' &&= (R(\theta_{23})R(\theta_{13}))^\dagger H R(\theta_{23})R(\theta_{13}) \nnb \\
   &&= U' H_0 (U')^\dagger +V',\label{H1}
\eea
where $U'=R(\theta_{12})R(\theta_{02})R(\theta_{01})R(\theta_{03})$ and
\bea
V'=\begin{pmatrix} 0 & 0 & 0 & 0 \cr
0 & V_e \cos^2\theta_{13} +V_n & 0 & V_e \sin\theta_{13}\cos\theta_{13} \cr
0 & 0 & V_n & 0 \cr
0& V_e \sin\theta_{13}\cos\theta_{13} &  0& V_e\sin^2\theta_{13}+V_n \cr
  \end{pmatrix}. \label{H1a}
\eea

Eq. (\ref{H1}) can be rewritten as
\bea
H'= \begin{pmatrix} {\widehat H}' & S \cr
S^\dagger & \frac{\Delta m^2_{31}}{2E}\cos^2\theta_{03}+\frac{\Delta m^2_{01}}{2E}\sin^2\theta_{03}+V_e\sin^2\theta_{13}+V_n \end{pmatrix}. \label{H2}
\eea
${\widehat H}'$ is a $3\times 3$ matrix
\bea
{\widehat H}'={\hat U}{\widehat H}'_0 {\hat U}^\dagger+{\widehat V}'.\label{H2b}
\eea
${\widehat H}'_0$, ${\widehat V}'$ and ${\hat U}$ are 
\bea
{\widehat H}'_0&& =\textrm{diag}\big\{ \frac{\Delta m^2_{01}}{2E}\cos^2\theta_{03}
+\frac{\Delta m^2_{31}}{2E}\sin^2\theta_{03}, 0,~
\frac{\Delta m^2_{21}}{2E} \big\} ,\label{H2c} \\
{\widehat V}'&&=\textrm{diag}\big\{0, V_e\cos^2\theta_{13}+V_n, V_n \big\}, \label{H2c-0} \\
{\hat U} &&= {\hat R}(\theta_{12}){\hat R}(\theta_{02}){\hat R}(\theta_{01}),\label{H2d}
\eea
where ${\hat R}(\theta_{ij})$ is a $3\times 3$ matrix with mixing angle appearing
at $i$ and $j$ entries, e.g.
\bea
{\hat R}(\theta_{01})=\begin{pmatrix} \cos\theta_{01} & \sin\theta_{01} & 0 \cr
-\sin\theta_{01} & \cos\theta_{01} & 0 \cr
0 & 0 & 1 \cr \end{pmatrix}, \label{H2e}
\eea
and
\bea
{\hat R}(\theta_{12})=\begin{pmatrix} 1 & 0 & 0 \cr
0 & \cos\theta_{12} & \sin\theta_{12} \cr
0 & -\sin\theta_{12} & \cos\theta_{12} \cr \end{pmatrix}. \label{H2f}
\eea
$S$ is a $3\times 1$ matrix
\bea
S={\hat U} \begin{pmatrix}
\frac{1}{2}\sin2\theta_{03}(\frac{\Delta m^2_{31}-\Delta m^2_{01}}{2E}) \cr
0 \cr
0 \cr
\end{pmatrix}
+\begin{pmatrix}0 \cr \frac{1}{2} \sin2\theta_{13}~ V_e \cr 0
\cr \end{pmatrix}, \label{H2a}
\eea

From (\ref{nuprime}) one can see that
$\nu'_e$ is mainly $\nu_e$ and has a small component of $\nu_\mu$ and $\nu_\tau$. It has
a probability $\cos^2\theta_{13}$ being $\nu_e$ and a probability $\sin^2\theta_{13}$
being $\nu_\mu$ and $\nu_\tau$. 
Disappearance of solar $\nu_e$ can be effectively
studied by examining the oscillation of $\nu'_e$ to other neutrinos~\cite{deHS}.

As is well known, $2-3$ resonance happens for energy $5-7$ GeV in the Earth and for
energy $< 1$ GeV, $|\Delta m^2_{31}/(2E)| \gg V_e$. In the
Earth, the neutron number density is roughly of the same order of
the electron number density. It is estimated that~\cite{lm}
\bea
 R_n =\left \{
 \begin{matrix} 0.024, & \textrm{~~~mantle} \cr
           0.146, & \textrm{core}
 \end{matrix}    \right. \label{RatioA}
\eea
where $R_n=(n_n-n_e)/n_e$.  So we can conclude that for energy $< 1$ GeV
we have $|\Delta m^2_{31}/(2E)| \gg |V_n|$. 
From this conclusion we can find from Eqs. (\ref{H2}) and (\ref{H2a})
that the correction to the probability
of the $\nu_s$ to $\nu'_e$ or $\nu'_\mu$ conversion
through $\nu'_\tau$ and vice versa, i.e.
\bea
\nu_s \leftrightarrows \nu'_\tau \leftrightarrows (\nu'_e,\nu'_\mu) , 
\eea
is suppressed by factor
\bea
\sin2\theta_{03} \frac{2EV_e}{\Delta m^2_{31}} \sin2\theta_{13}
~~\textrm{or}~~ \sin^2 2\theta_{03}.
\eea
For $\sin^22\theta_{13}\approx 0.09$
~\cite{theta13,theta13-2,theta13-3} and $\sin^2 2\theta_{03} \lsim 0.001$,
this amplitude is maximally of order $10^{-3}$ for $E < 1$ GeV.
Hence the effects of $\nu_s-(\nu'_e,\nu'_\mu)$ oscillation through $\nu'_\tau$
is sub-leading and can be neglected. 

\begin{figure}[tb]
\begin{center}
\begin{tabular}{cc}
\includegraphics[scale=1,width=8cm]{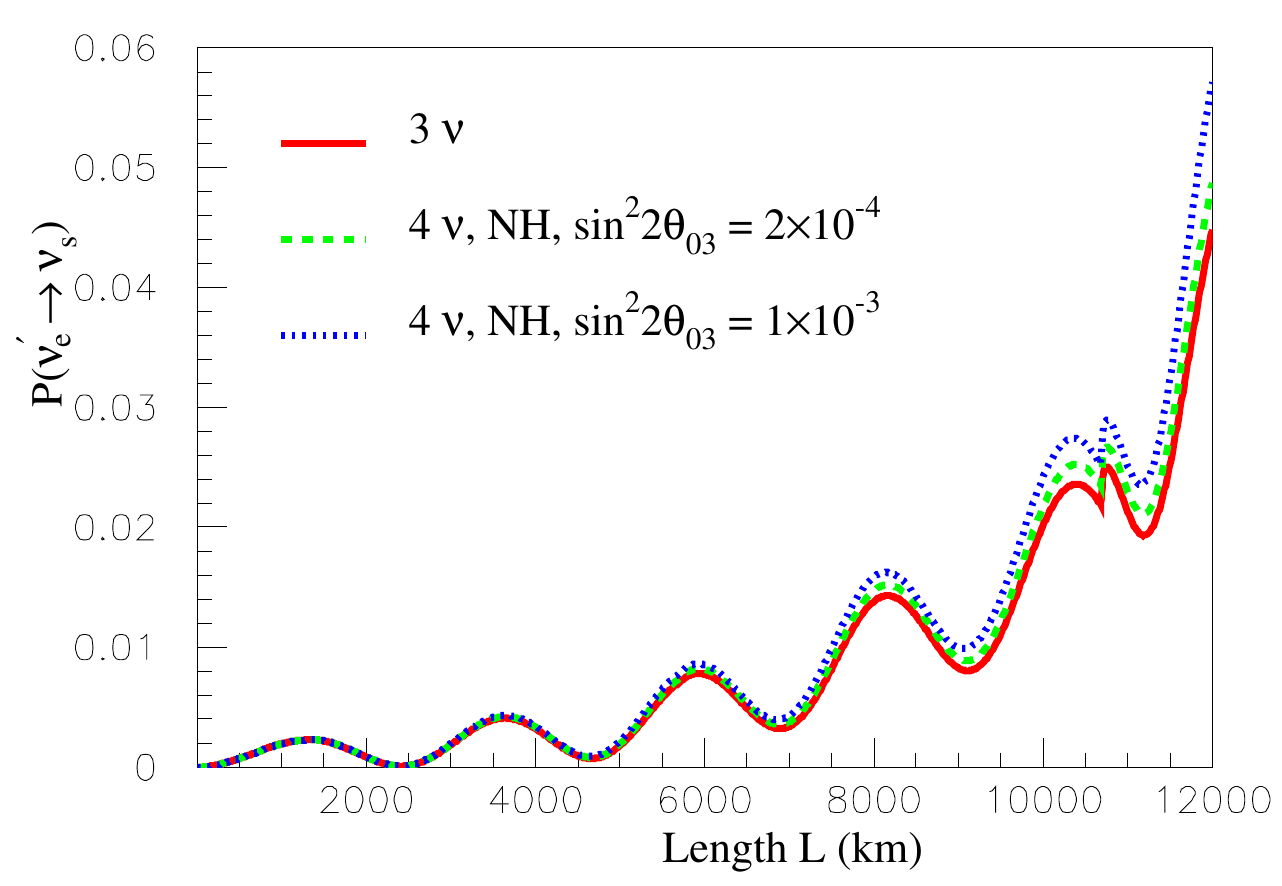}
\includegraphics[scale=1,width=8cm]{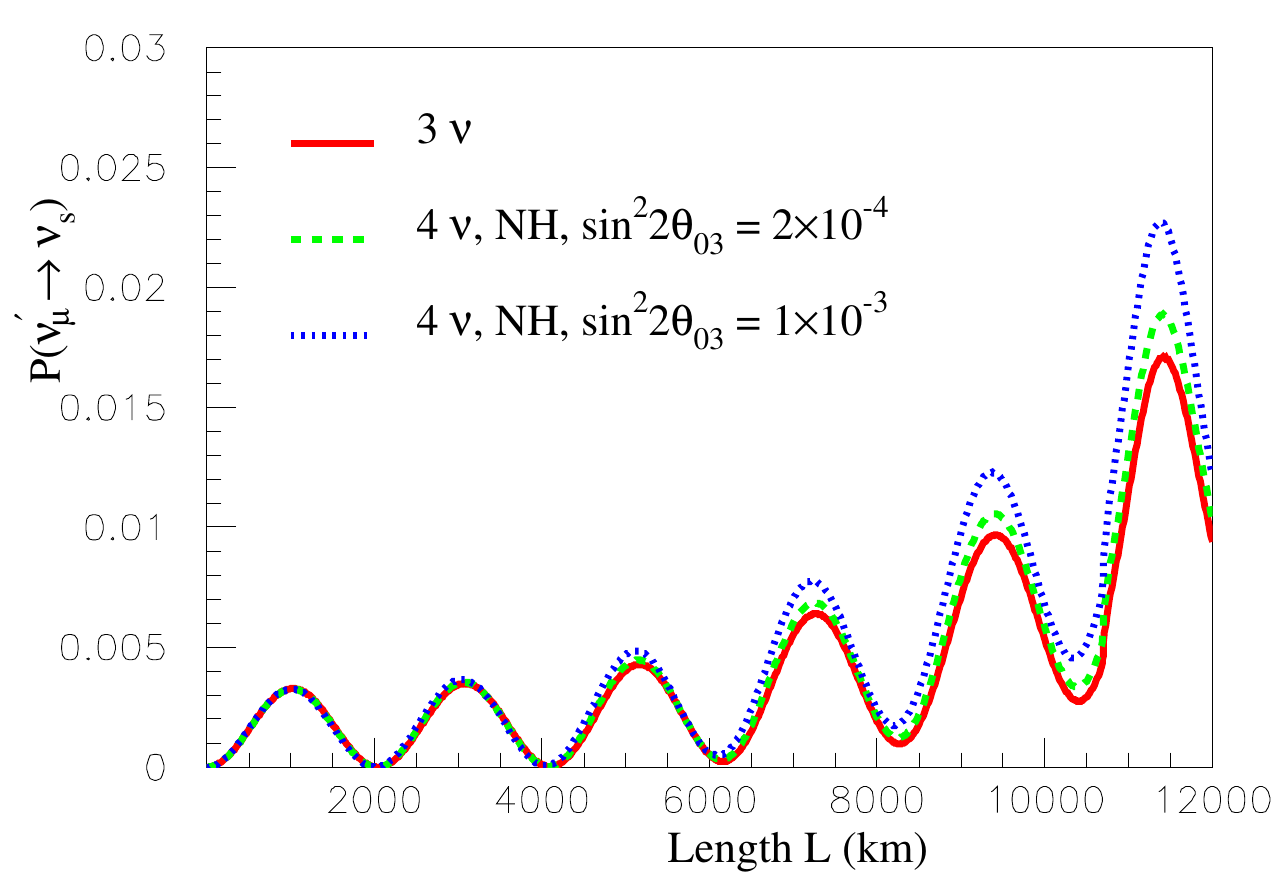}
\\
\includegraphics[scale=1,width=8cm]{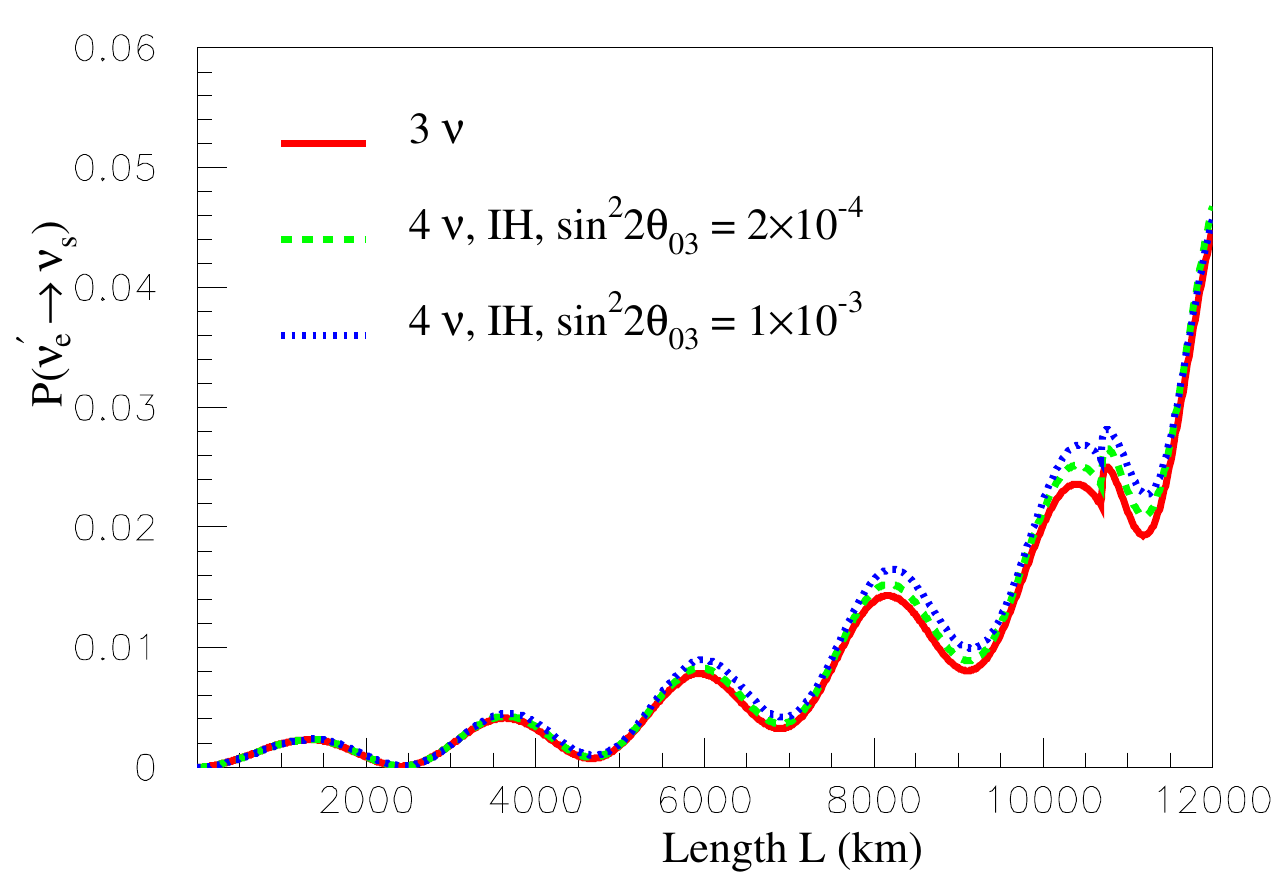}
\includegraphics[scale=1,width=8cm]{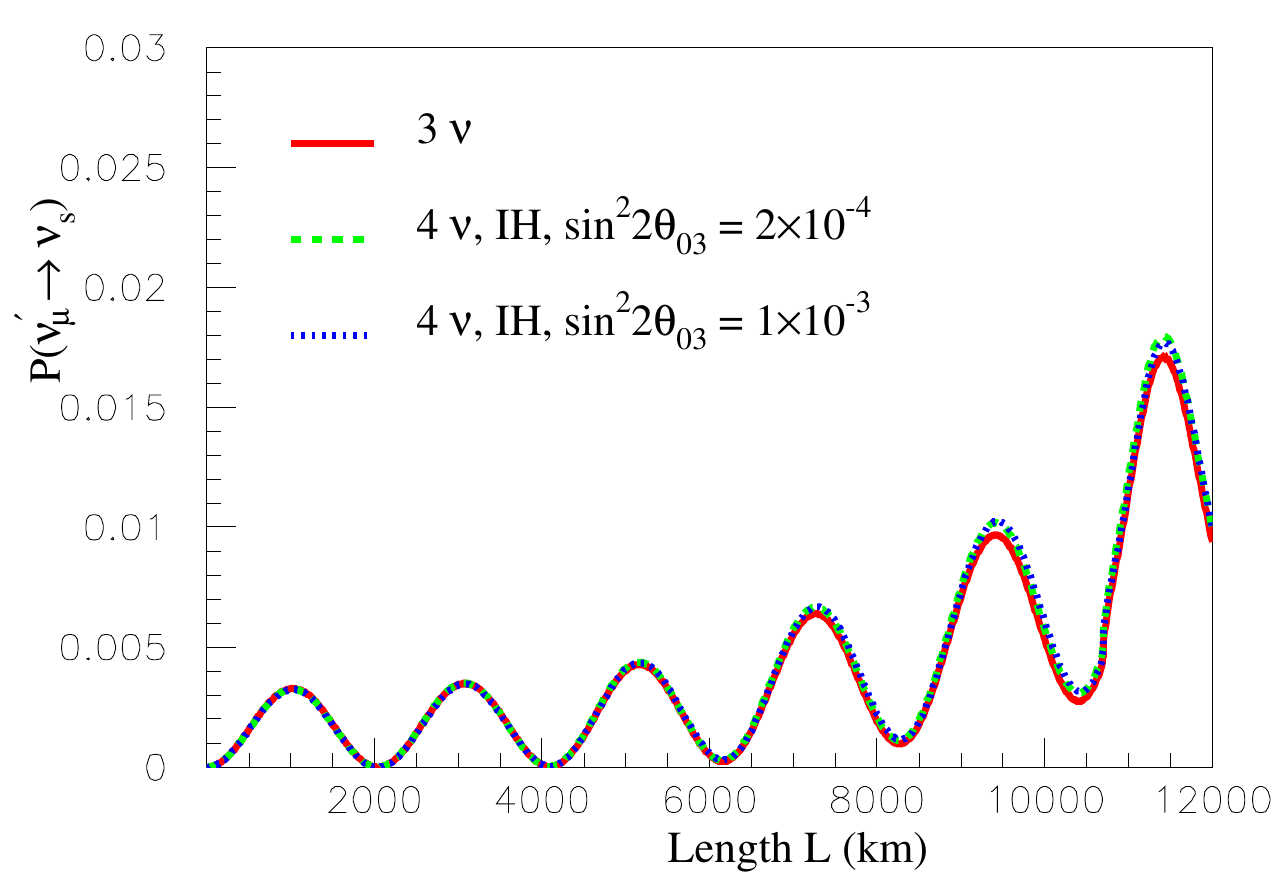}
\end{tabular}
\end{center}
\caption{$\nu'_e-\nu_s$ and $\nu'_\mu-\nu_s$ conversion probability
in three neutrino framework
v.s. in four neutrino framework. $E=60$ MeV,
$\Delta {\widehat m}^2_{01}=0.5\times 10^{-5}$ eV$^2$,
$\sin^22\theta_{02}=0.005$, $\sin^22\theta_{01}=0$.}
\label{figure1}
\end{figure}

In Fig. \ref{figure1} we give the plots of $\nu_s-(\nu'_e,\nu'_\mu)$ conversion
calculated in the $4\nu$ framework using Hamiltonian (\ref{H2}) 
and in the $3\nu$ framework using ${\widehat H}'$ in
Hamiltonian (\ref{H2}) respectively. 
Relevant neutrino parameters in our calculation are~\cite{RPP}
\bea
& \Delta m^2_{21}=(7.50\pm 0.20) \times 10^{-5} ~\textrm{eV}^2,~~
|\Delta m^2_{32}|=(2.32^{+0.12}_{-0.08})\times 10^{-3} ~\textrm{eV}^2, \label{neuparam0} \\
& \sin^2 2\theta_{12}=0.857\pm 0.024, ~~\sin^2 2\theta_{23}>0.95. \label{neuparam1}
\eea
Soon after the discovery of a not small $\theta_{13}$ by Daya-Bay collaboration
\cite{theta13}, confirmed by RENO experiment~\cite{theta13-2}, a precise
measurement of $\theta_{13}$ has been achieved by Daya-Bay experiment~\cite{theta13-3}:
\bea
\sin^22\theta_{13}=0.089\pm 0.010\pm 0.005 . \label{neuparam2}
\eea 
We use $\sin^22\theta_{23}=1$ in our calculation. For other parameters
we use the central values in Eqs. (\ref{neuparam0}), 
(\ref{neuparam1}) and (\ref{neuparam2}).
Results in Fig. \ref{figure1}
have been shown both for the case of normal hierarchy(NH)
and for the case of inverted hierarchy(IH).

In Fig. \ref{figure1} we can see that the result calculated in $3\nu$ framework
agrees with that calculated in $4\nu$ framework in the case
that $\sin^2 2\theta_{03}$ is small. Actually, for $\theta_{03}=0$
two lines for $4\nu$ and $3\nu$ overlap. 
We see that a non-zero but small $\theta_{03}$ can
not change the conversion qualitatively.
This agrees with our discussion presented above.

In the following we will discuss effects of $\theta_{01}$ and 
$\theta_{02}$ in neutrino oscillation and will set $\theta_{03}=0$.
So we can study the oscillation
of $\nu_s-(\nu'_e,\nu'_\mu)$ effectively in a $3\nu$ framework
\bea
i \frac{d}{d t} {\widehat \nu}'={\widehat H} {\widehat \nu'}, \label{H3}
\eea
where ${\widehat \nu}'=(\nu_s,\nu'_e,\nu'_\mu)^T$.
${\widehat H}$ is given by
\bea
{\widehat H}={\hat U} {\widehat H}_0 {\hat U}^\dagger+{\widehat V}, \label{H3-0}
\eea
with ${\widehat H}_0$ obtained from (\ref{H2c}) as
\bea
{\widehat H}_0 =\textrm{diag}\big\{ \frac{\Delta {\widehat m}^2_{01}}{2E},
0, \frac{\Delta m^2_{21}}{2E} \big\}, \label{H3a}
\eea
where
\bea
\Delta {\widehat m}^2_{01}=\Delta m^2_{01} \cos^2\theta_{03}+\Delta m^2_{31} \sin^2\theta_{03},
\label{H3b}
\eea
and 
\bea
{\widehat V}=\textrm{diag}\big\{0, V_e+V_n,V_n \big\}. \label{H3c}
\eea
(\ref{H3c}) is obtained from (\ref{H2c-0}) by approximating $\cos^2\theta_{13}=1$.
Since $\sin^22\theta_{13}\approx 0.09$ and 
$\sin^2\theta_{13}\approx 0.022$ this is a valid approximation.
$\Delta {\widehat m}^2_{01}$ is basically the parameter governing the
solar $\nu_e$ disappearance to $\nu_s$ discussed in ~\cite{deHS}.
In the following we will will set $\theta_{03}=0$.
Hence we will not differentiate between $\Delta m^2_{01}$ and
$\Delta {\widehat m}^2_{01}$ in the following of the present
article.

\begin{figure}[tb]
\begin{center}
\begin{tabular}{cc}
\includegraphics[scale=1,width=8cm]{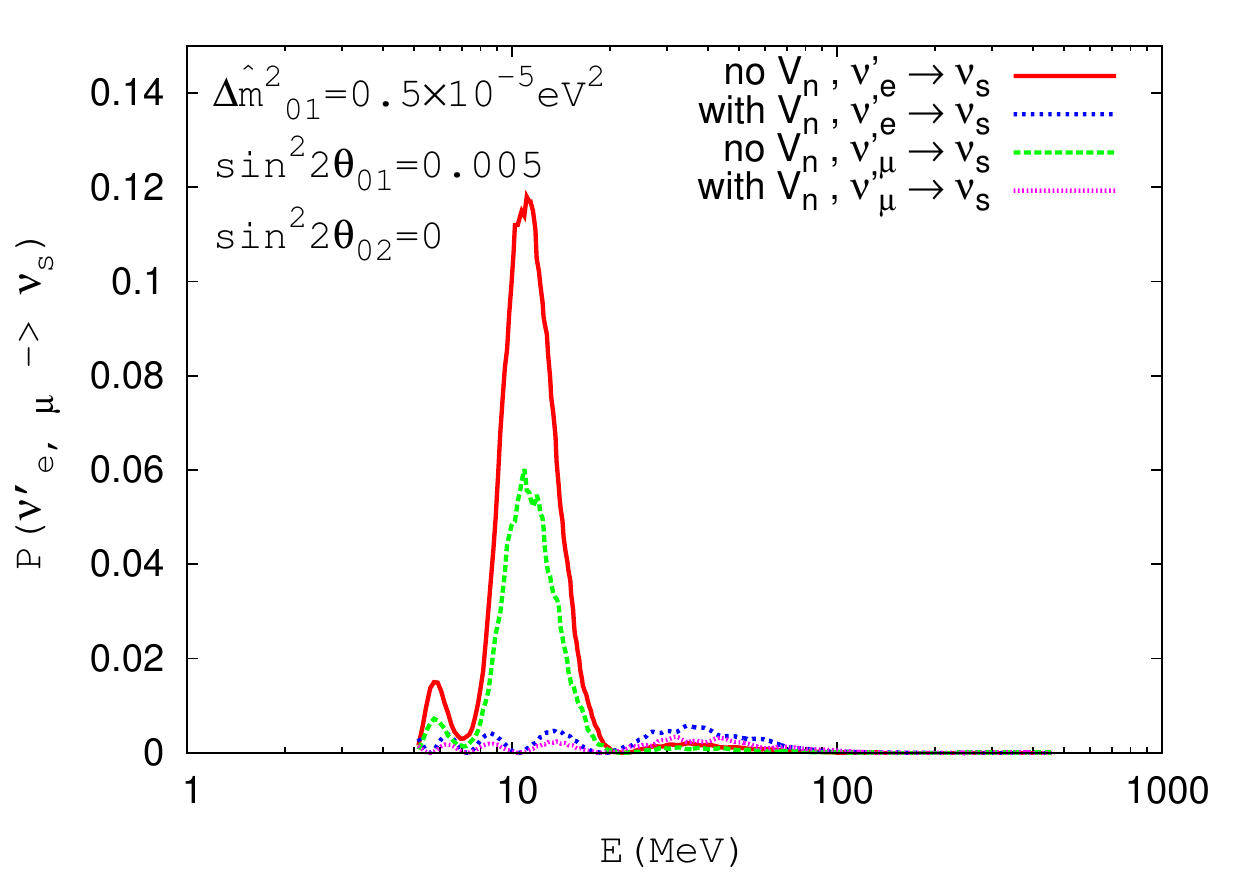}
\includegraphics[scale=1,width=8cm]{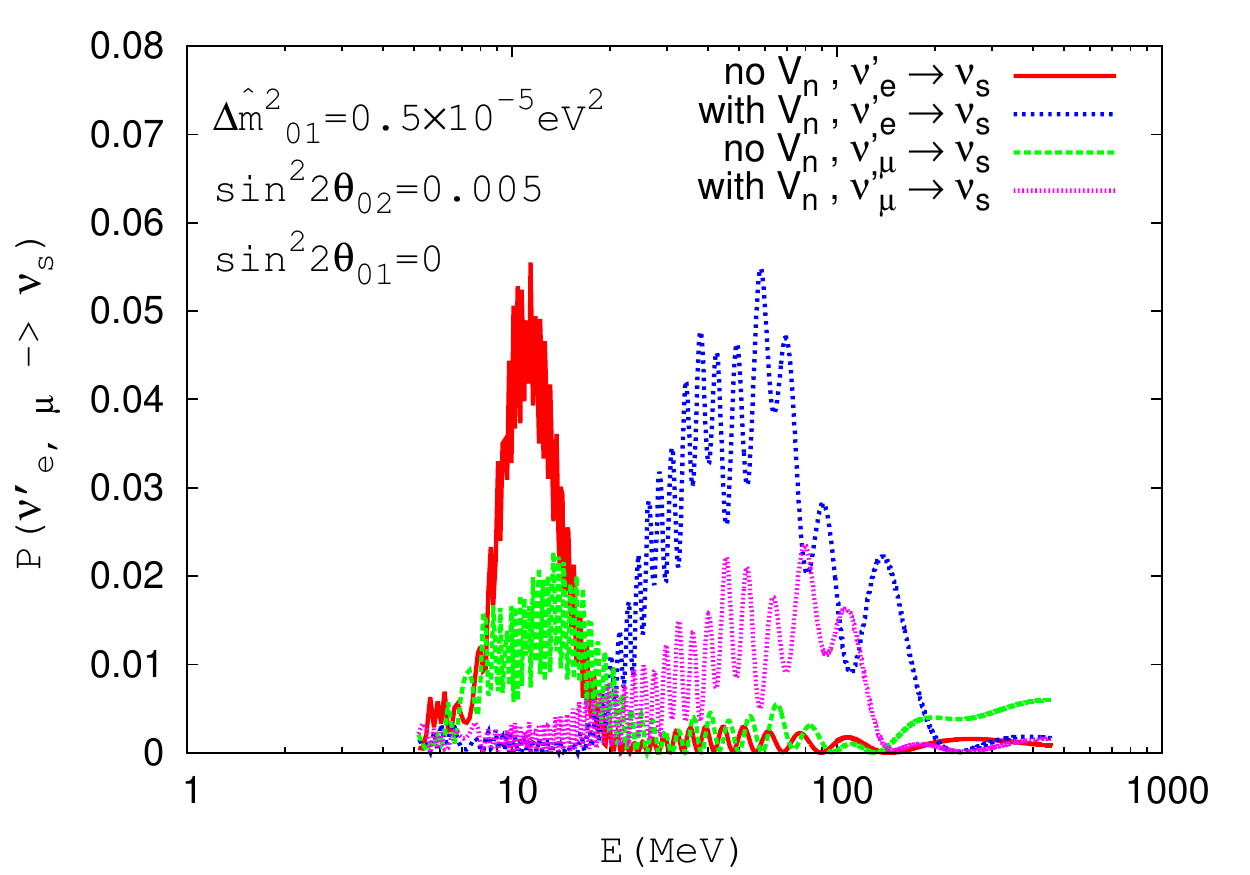}
\\
\includegraphics[scale=1,width=8cm]{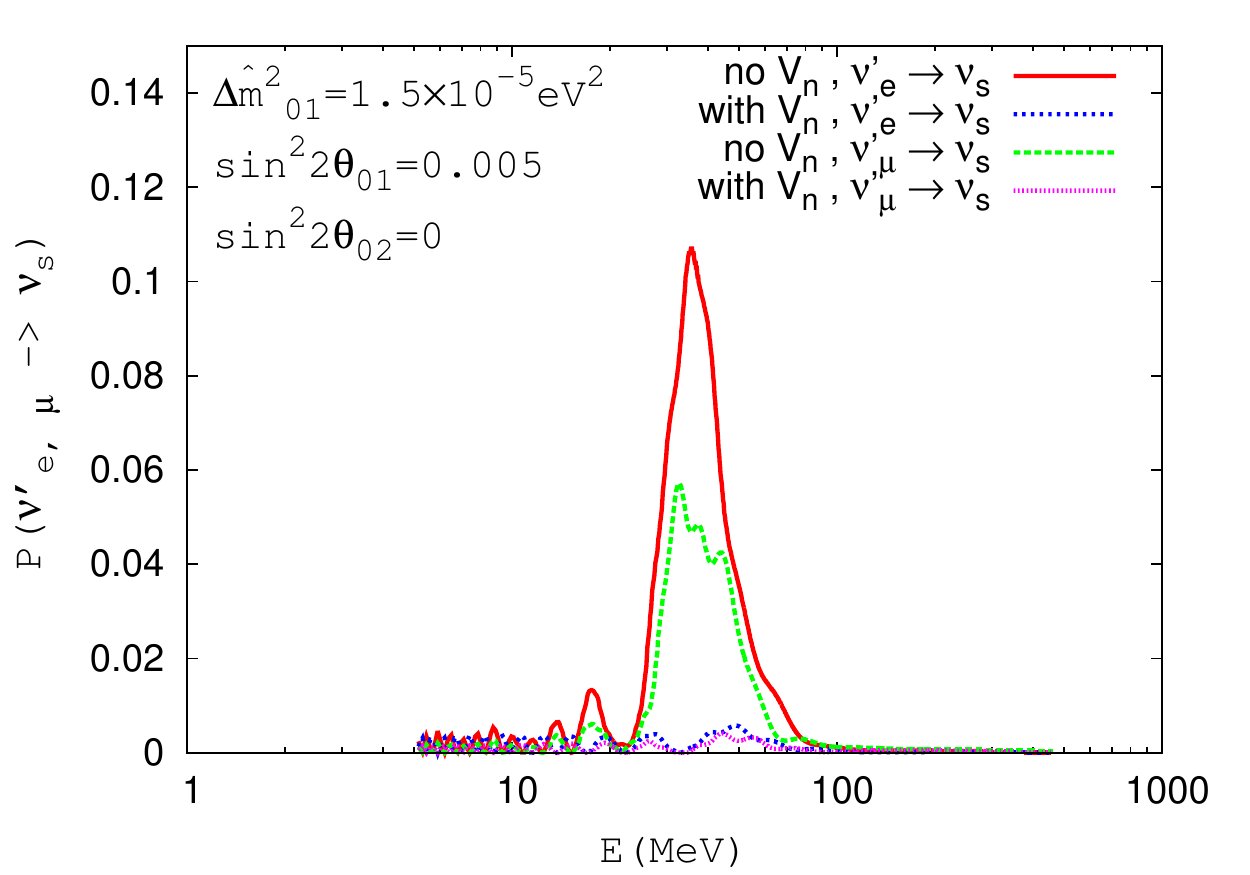}
\includegraphics[scale=1,width=8cm]{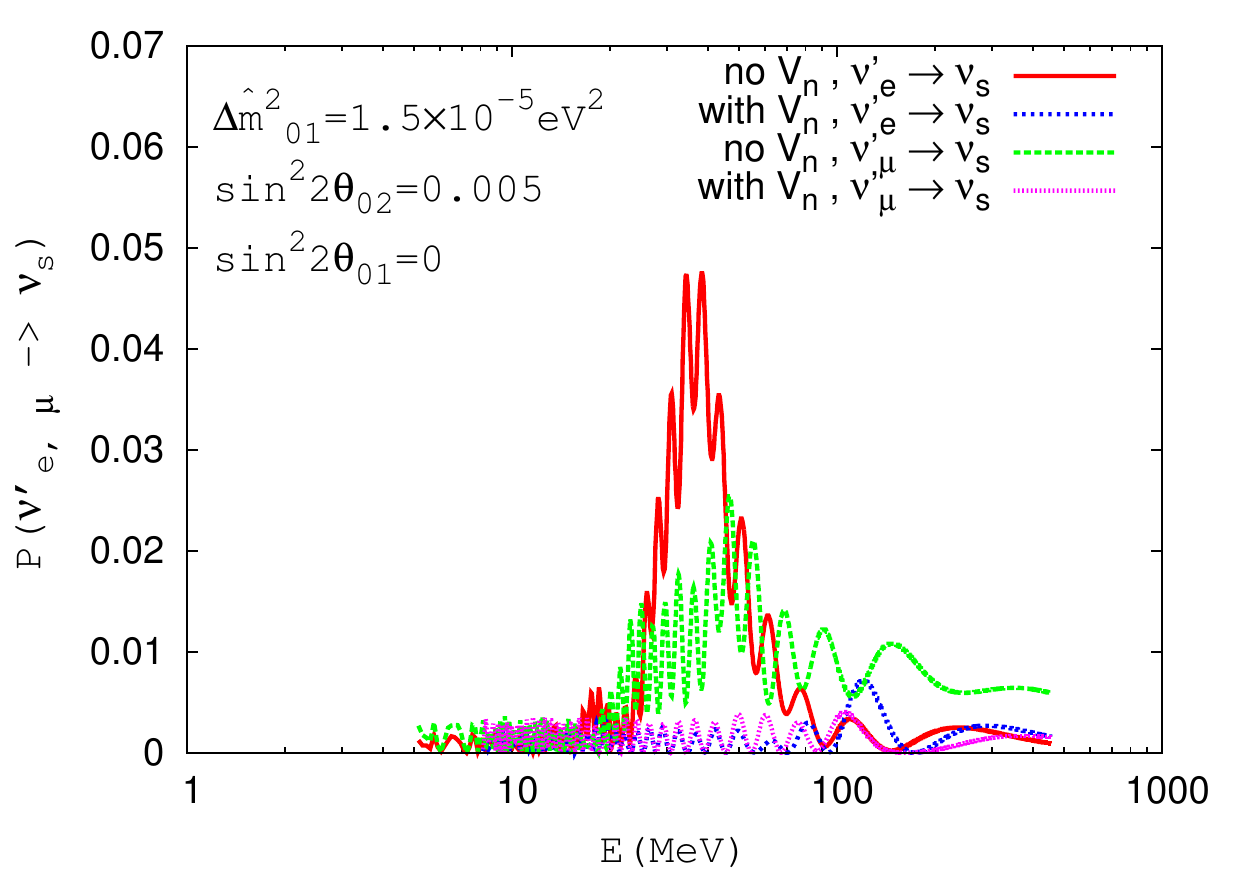}
\end{tabular}
\end{center}
\caption{$\nu'_e-\nu_s$ and $\nu'_\mu-\nu_s$ conversion probability
versus energy in three neutrino framework both for $V_n$ included and
for $V_n$ switched off. $L=12000$ km.
}
\label{figure0}
\end{figure}

Fig. \ref{figure1} is given with the effect of $V_n$ included and
has been shown to a particular energy that $\nu_s-\nu'_e$ conversion 
is close to maximal. 
In Fig. \ref{figure0} we give plots of the conversion probability
of $\nu'_{e, \mu} \to \nu_s$ versus energy both for $V_n$ included and for
$V_n$ switched off. For the case with the effect of $V_n$ included
the neutron number density is calculated using (\ref{RatioA}). 
We can see that for the case with the  effect of $V_n$ included 
the probability of $\nu'_e \to \nu_s$ conversion is maximally
around $5\%$ for $\sin^22\theta_{02}=0.005$. 
For smaller $\sin^22\theta_{02}$ the amplitude of the
conversion probability is smaller.
This enhancement of conversion probability
happens for $\Delta {\widehat m}^2_{01}$ around
$0.5\times 10^{-5}$ eV$^2$ and for energy around 60 MeV. 
For $\Delta {\widehat m}^2_{01}=1.5\times 10^{-5}$ eV$^2$, the conversion probability
is maximally around $10^{-3}$ and the resonant conversion disappears.

In contrast, we can see in Fig. \ref{figure0} that there are indeed 
much stronger resonant $\nu_s-(\nu'_e,\nu'_\mu)$ conversions
when $V_n$ is switched off.
In particular, there is a resonance conversion for energy around 10 MeV
when $\Delta {\widehat m}^2_{01}$ is around $0.5\times 10^{-5}$ eV$^2$.
This is consistent with our expectation.
Fortunately, this resonance disappears after including effect of $V_n$.
Although there is still a resonant enhancement of $\nu'_{e,\mu}\to \nu_s$
conversion for $\Delta {\widehat m}^2_{01}=0.5\times 10^{-5}$eV$^2$
and $\sin^2 2\theta_{02}=0.005$ when effect of $V_n$ included, 
the conversion probability can only reach about
$5 \%$ and furthermore it appears at energy around 60 MeV which is
well beyond the solar neutrino and supernovae neutrino spectrum.
For larger $\Delta {\widehat m}^2_{01}$ this enhancement disappears
completely, as can be seen in the plot for 
$\Delta {\widehat m}^2_{01}=1.5\times 10^{-5}$eV$^2$ in Fig. \ref{figure0}.
More details about the variation with respect to $\Delta {\widehat m}^2_{01}$
will be presented in the next section.

This is a very interesting observation. If the resonant conversion
happens as in the case with $V_n$ switched off, active neutrinos 
of energy around 10 to 60 MeV 
can oscillate to this super-light sterile. In particular,
this could lead to spectrum distortion of high energy solar boron neutrino
and the flux of low energy atmospheric active neutrinos should have a dip at this
energy range. The disappearance or the suppression of
the resonant conversion in the case with $V_n$ included shows that
the scenario of super-light sterile neutrino can pass through
the possible constraints from solar and atmospheric neutrino experiments.

We note that this phenomenon does not happen for the conversion of
neutrino in the Sun since the neutron number density is small
in the Sun and the effect of $V_n$ is indeed switched off. 
On the other hand, the
neutron number density is of the same order of the magnitude of
the electron number density in the Earth and the effect of $V_n$ can play important
role. In the next section we show in detail how the level crossing
and the resonance disappear
in the Earth matter.

{\bf Level crossing and flavor conversion of the super-light sterile neutrino}
\begin{figure}[tb]
\begin{center}
\begin{tabular}{cc}
\includegraphics[scale=1,width=7.5cm]{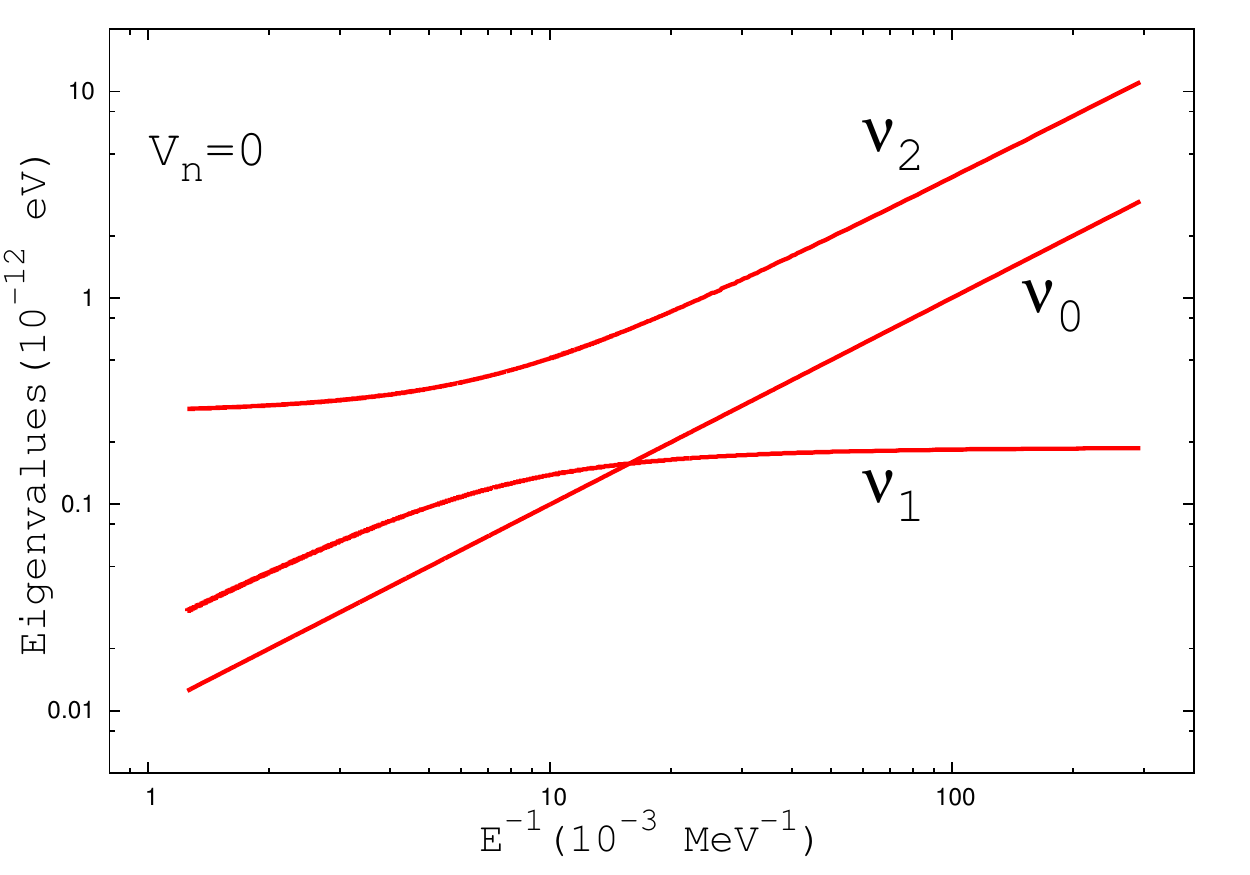}
\includegraphics[scale=1,width=7.5cm]{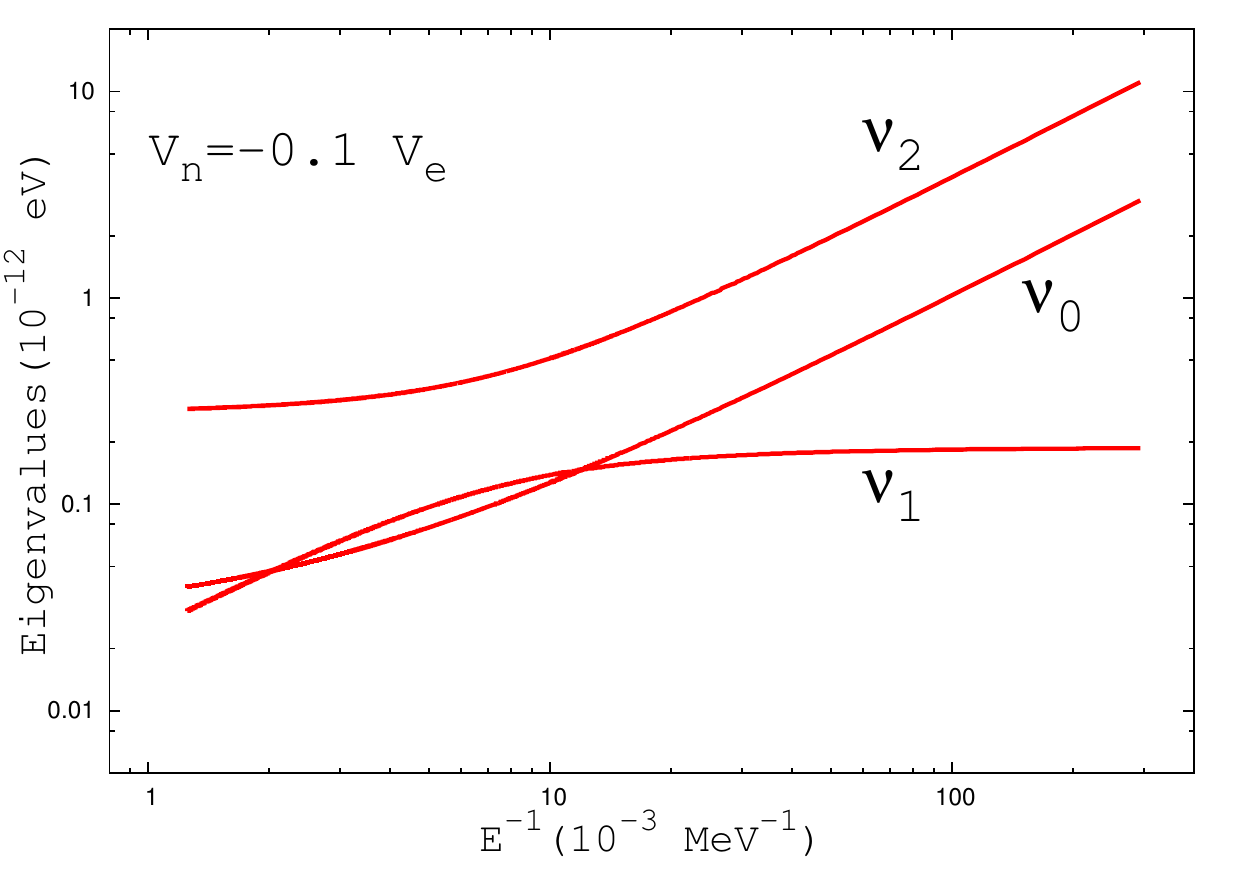}
\\
\includegraphics[scale=1,width=7.5cm]{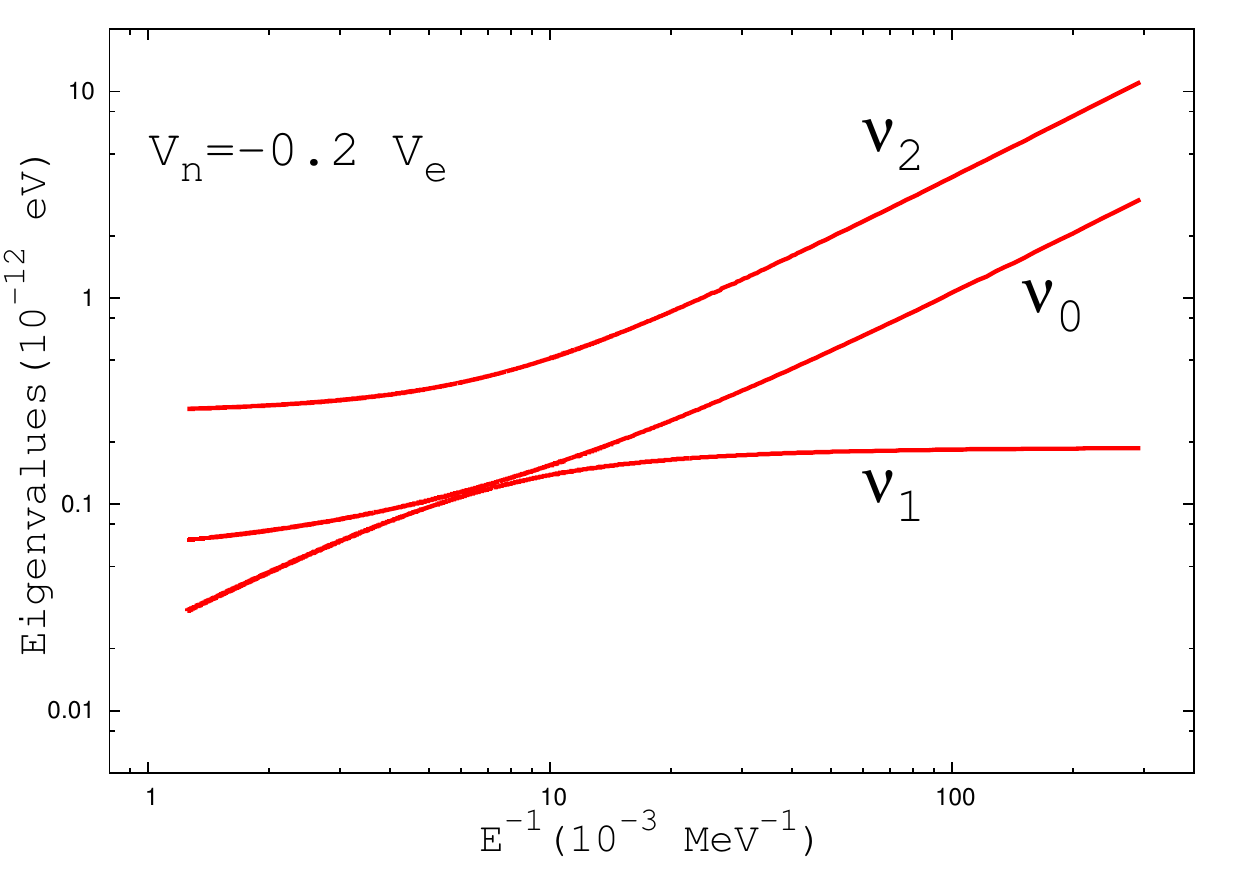}
\includegraphics[scale=1,width=7.5cm]{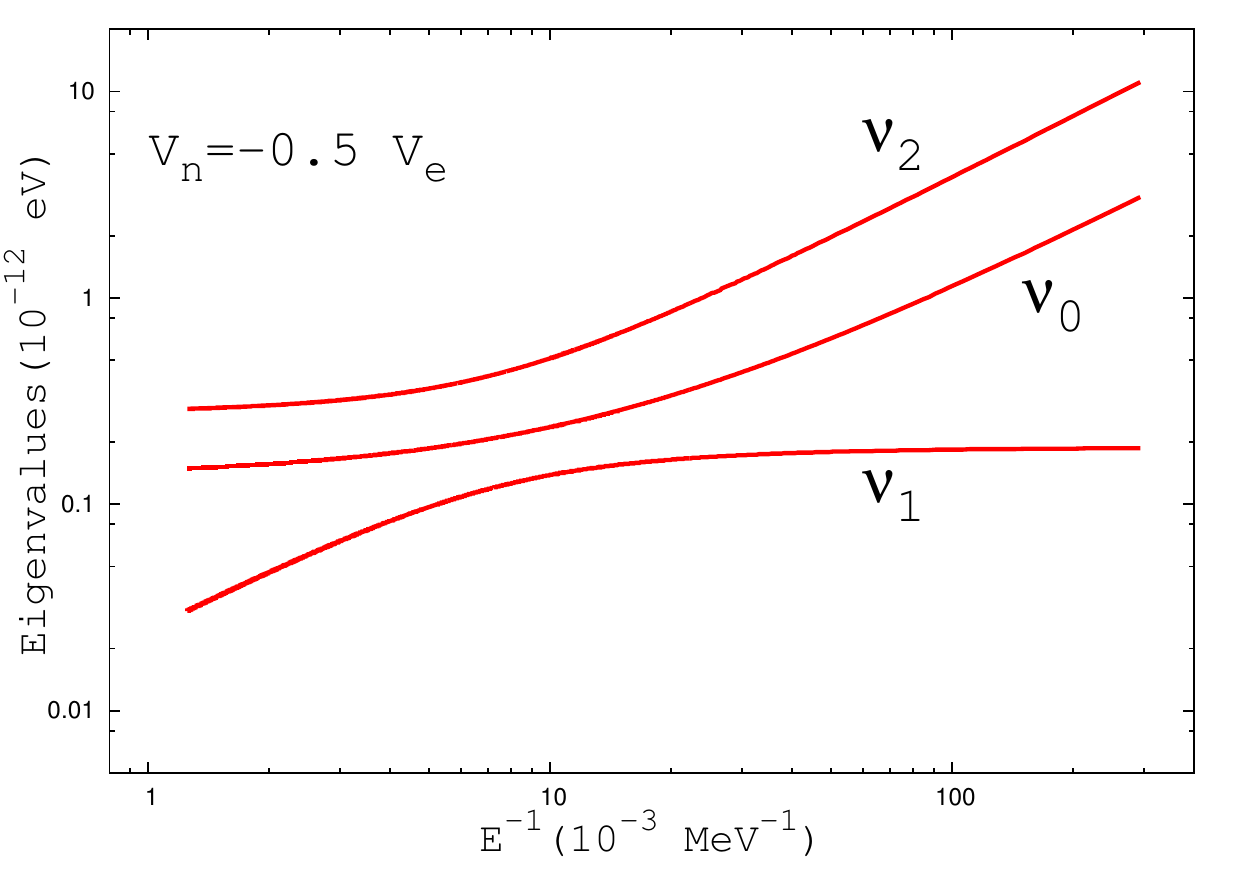}
\end{tabular}
\end{center}
\caption{Energy levels $\nu_0$, $\nu_1$ and $\nu_2$ in three neutrino
framework. Three lines for $\nu_0$, $\nu_1$ and $\nu_2$ are $E_0-V_n$,
$E_1-V_n$ and $E_2-V_n$ separately. $\theta_{01}=\theta_{02}=0$,
$\Delta {\widehat m}^2_{01}=2\times 10^{-5}$ eV$^2$,
$\Delta m^2_{21}=7.5 \times 10^{-5}$ eV$^2$, $\sin^2 2\theta_{12}=0.857$,
$V_e=2.74\times 10^{-13}$ eV.}
\label{figure2}
\end{figure}

In this section we study the effect of $V_n$ on the level crossing and
the resonant conversion of
the super-light sterile neutrino $\nu_s$ with active neutrinos $\nu'_e$
and $\nu'_\mu$ in the Earth. We will work in the $3\nu$ framework
as given in Eqs. (\ref{H3}), (\ref{H3-0}), (\ref{H3a}), (\ref{H3b})
and (\ref{H3c}).

A MSW resonance of flavor conversion happens when two energy levels become close
to each other. If a resonance happens a small mixing in vacuum can effectively
lead to large flavor conversion in matter. This happens in particular
for the case that there is a crossing of two energy levels of the
Hamiltonian. To examine whether there is a level crossing or
a resonance it's crucial to check whether two eigenvalues of the
Hamiltonian ${\widehat H}$ become close to each other. 
As will be shown below, it's sufficient to set small parameters
$\theta_{01}$ and $\theta_{02}$ to zero when studying the energy levels 
of ${\widehat H}$. For convenience
we will set $\theta_{01}=\theta_{02}=0$ and discuss the energy levels
of the following Hamiltonian
\bea
{\widehat H}=\begin{pmatrix} \frac{\Delta {\widehat m}^2_{01}}{2E} & 0 & 0 \cr
0 & \frac{\Delta m^2_{21}}{2E}\sin^2\theta_{12}+ V_e+V_n & 
\frac{\Delta m^2_{21}}{2E}\sin\theta_{12}\cos\theta_{12} \cr
0 & \frac{\Delta m^2_{21}}{2E}\sin\theta_{12} \cos\theta_{12} & 
\frac{\Delta m^2_{21}}{2E}\cos^2\theta_{12}+V_n \cr \end{pmatrix}. \label{H4}
\eea
(\ref{H4}) is obtained from (\ref{H3-0}) by setting $\theta_{01}=\theta_{02}=0$
in ${\hat U}$.

The convenience of using (\ref{H4}) is clear by noting that the
three eigenvalues of (\ref{H4}) can be easily found as
\bea
E_0&& =\frac{\Delta {\widehat m}^2_{01}}{2E},\label{E0} \\
E_1&& =\frac{1}{2}\bigg[ \frac{\Delta m^2_{21}}{2E}
+V_e-\sqrt{(\frac{\Delta m^2_{21}}{2E}\cos2\theta_{12}-V_e)^2+(\frac{\Delta m^2_{21}}{2E}
\sin2\theta_{12})^2 }\bigg ]+ V_n,\label{E1} \\
E_1&& =\frac{1}{2}\bigg[ \frac{\Delta m^2_{21}}{2E}
+V_e+\sqrt{(\frac{\Delta m^2_{21}}{2E}\cos2\theta_{12}-V_e)^2+(\frac{\Delta m^2_{21}}{2E}
\sin2\theta_{12})^2 }\bigg] +V_n.\label{E2} 
\eea
$E_0$, $E_1$, $E_2$ are eigenvalues of (\ref{H4}) corresponding to
neutrinos in mass base $\nu_0$, $\nu_1$ and $\nu_2$ separately.
To illustrate qualitatively the effect of the Earth matter
on the oscillation of super-light sterile neutrino
$\nu_s$ with active neutrinos $\nu'_e$ and $\nu'_\mu$, it's 
convenient to compute $E_1$ and $E_2$ 
using a trajectory dependent averaged potential~\cite{Liao}
\bea
{\bar V}_e = \frac{1}{L} \int^L_0 ~dx~V_e(x), \label{poten}
\eea
where $L$ is the length of the neutrino trajectory in the Earth.
For baseline longer than 6000 km, ${\bar V}_e$ varies from 
$1.6\times 10^{-13}$ eV to about $2.7\times 10^{-13}$ eV.
As will be detailed in the next section a formulation
using the average potential (\ref{poten}) gives
a pretty good description of the oscillation of super-light
sterile neutrino with active neutrinos.

In Fig. \ref{figure2} we give plots for $E_i-V_n$ and compare different
cases with various $V_n$. We can see that for $V_n=0$ there is indeed a
level crossing of $E_0$ and $E_1$ at energy around $50-60$ MeV
when $\Delta {\widehat m}^2_{01}=2\times 10^{-5}$ eV$^2$ and there 
should be a resonant conversion associated with it, as has been 
shown in Fig. \ref{figure0} for $\Delta {\widehat m}^2_{01}=1.5\times 10^{-5}$ eV$^2$.
Increasing the magnitude
of $V_n$ first gives rise to a second point of level crossing, as can be seen in
the plot with $V_n=-0.1 V_e$. But for $V_n=-0.5 V_e$ there is no
crossing point of two lines of $E_0$ and $E_1$ and the resonance disappears.
We note that this is exactly the situation we have in the Earth
matter. As can be seen in Eq. (\ref{RatioA}), the neutron number
density roughly equals to the electron number density in the Earth and hence
$V_n \approx -0.5 V_e$.

\begin{figure}[tb]
\begin{center}
\begin{tabular}{cc}
\includegraphics[scale=1,width=8.2cm]{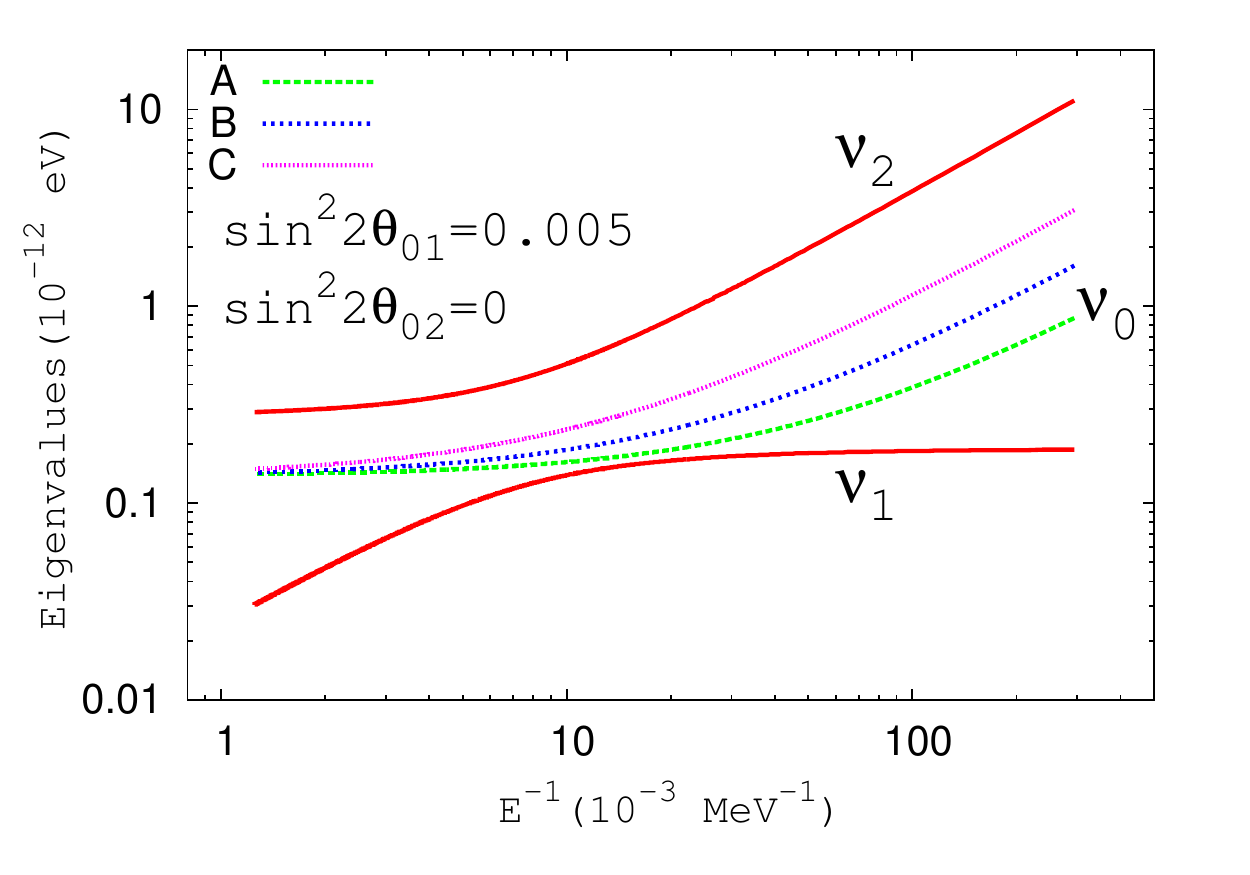}
\includegraphics[scale=1,width=8.2cm]{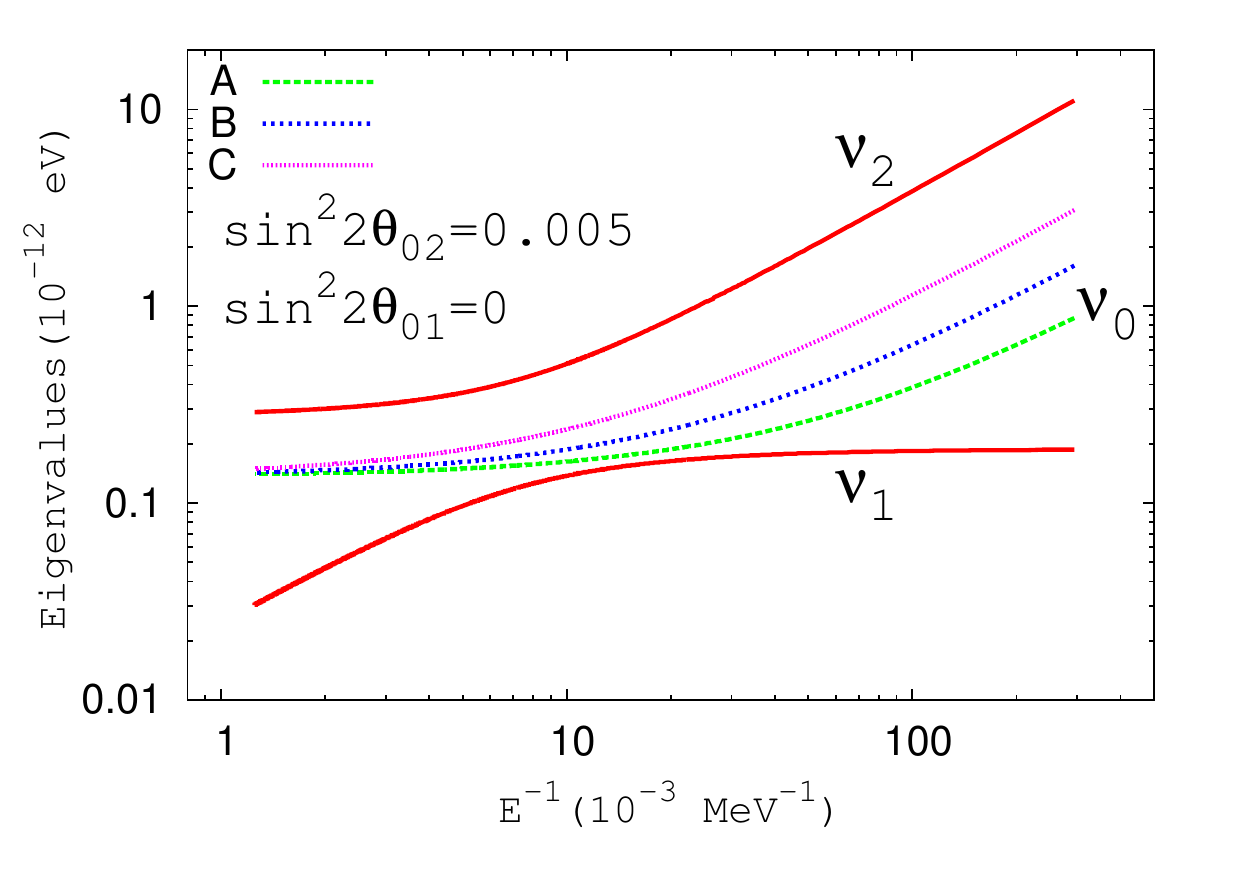}
\end{tabular}
\end{center}
\caption{$E_0-V_n$, $E_1-V_n$ and $E_2-V_n$ for
 $\nu_0$, $\nu_1$ and $\nu_2$ in three neutrino
framework. $V_n=-0.5 V_e$ and $V_e=2.74 \times 10^{-13}$ eV. Lines A, B and 
C correspond to cases with $\Delta {\widehat m}^2_{01}=0.5 \times 10^{-5}$ eV$^{2}$,
$\Delta {\widehat m}^2_{01}=1.0 \times 10^{-5}$ eV$^{2}$ and 
$\Delta {\widehat m}^2_{01}=2.0 \times 10^{-5}$ eV$^{2}$ separately.}
\label{figure3}
\end{figure}

For a non-zero but small $\theta_{01}$ or $\theta_{02}$ one can give
similar plots for $E_0$, $E_1$ and $E_2$ and there are no visible differences
compared to the plots in Fig. \ref{figure2}. So the above
discussion presented for $\theta_{01}=\theta_{02}=0$ can be applied to the case
with non-zero but small $\theta_{01}$ or $\theta_{02}$
and the reason of the absence of resonance for a relatively large value
of $\Delta{\widehat m}^2_{01}$ is clear according to discussion presented above.
In Fig. \ref{figure3} we give plots of $E_i-V_n$ for $\theta_{01,02}\neq 0$.
$E_i$ are obtained from diagonalizing (\ref{H3-0}) numerically.
$V_n/V_e$ is fixed in Fig. \ref{figure3} while $\Delta {\widehat m}^2_{01}$
varies. In Earth matter 
$V_e/V_n$ can vary from $-0.5012$ to $-0.573$. As can be seen in
Fig. \ref{figure2}, the larger the $|V_e/V_n|$ is,
the farther away from the level crossing the two energy levels. So we choose
$V_e=-0.5 V_n$ in Fig. \ref{figure3}.
We can see that for $\Delta {\widehat m}^2_{01}=(0.5-2)\times 10^{-5}$ eV$^2$
the line of $E_0$ is always in-between the two lines of $E_1$ and $E_2$.
The only possible case for a resonant conversion to happen is when
$\Delta m^2_{01}\approx 0.5\times 10^{-5}$ eV$^2$. In this case
$E_0$ and $E_1$ come close to each other although there is no crossing.
However, even in this case the $\nu'_{e,\mu}\to\nu_s$ conversion probability
is maximally around $5\%$ when $\sin^22\theta_{02}=0.005$, as has been
shown in Fig. \ref{figure0}.

\begin{figure}[tb]
\begin{center}
\begin{tabular}{cc}
\includegraphics[scale=1,width=8.2cm]{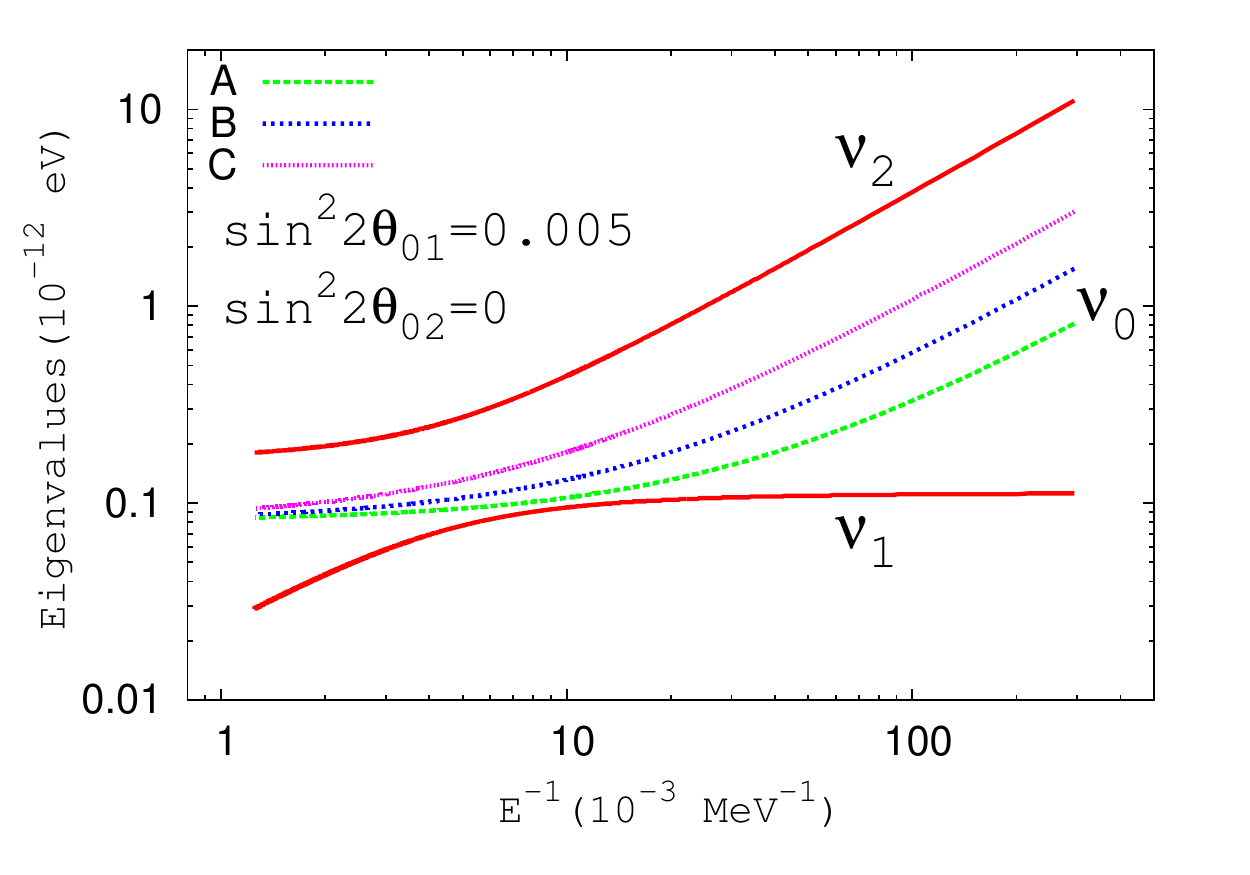}
\includegraphics[scale=1,width=8.2cm]{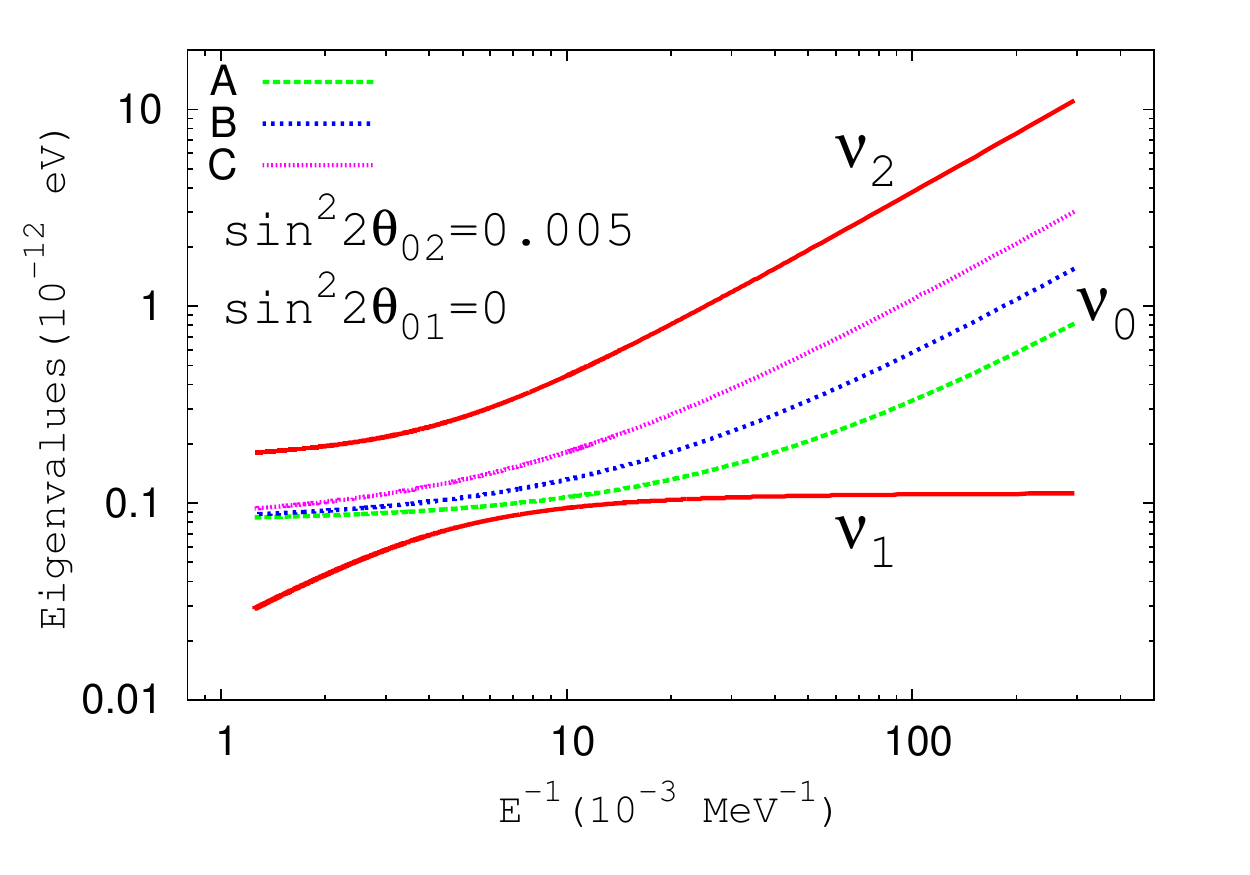}
\end{tabular}
\end{center}
\caption{$E_0-V_n$, $E_1-V_n$ and $E_2-V_n$ for
 $\nu_0$, $\nu_1$ and $\nu_2$ in three neutrino
framework. $V_n=-0.5 V_e$ and $V_e=1.63 \times 10^{-13}$ eV. Lines A, B and 
C correspond to cases with $\Delta {\widehat m}^2_{01}=0.5 \times 10^{-5}$ eV$^{2}$,
$\Delta {\widehat m}^2_{01}=1.0 \times 10^{-5}$ eV$^{2}$ and 
$\Delta {\widehat m}^2_{01}=2.0 \times 10^{-5}$ eV$^{2}$ separately.}
\label{figure4}
\end{figure}

In Fig. \ref{figure4} we give plots of $E_i-V_n$ similar to
Fig. \ref{figure3} but with different $V_e$. We can see that the
situation is very similar to that in Fig. \ref{figure3} and
there are no level crossing of $E_0$ and $E_1$ energy levels.
From the above discussion we can conclude that due to the
presence of $V_n$ in Earth matter the level crossing expected
for $E_0$ and $E_1$ disappears.

\begin{figure}[tb]
\begin{center}
\begin{tabular}{cc}
\includegraphics[scale=1,width=8.2cm]{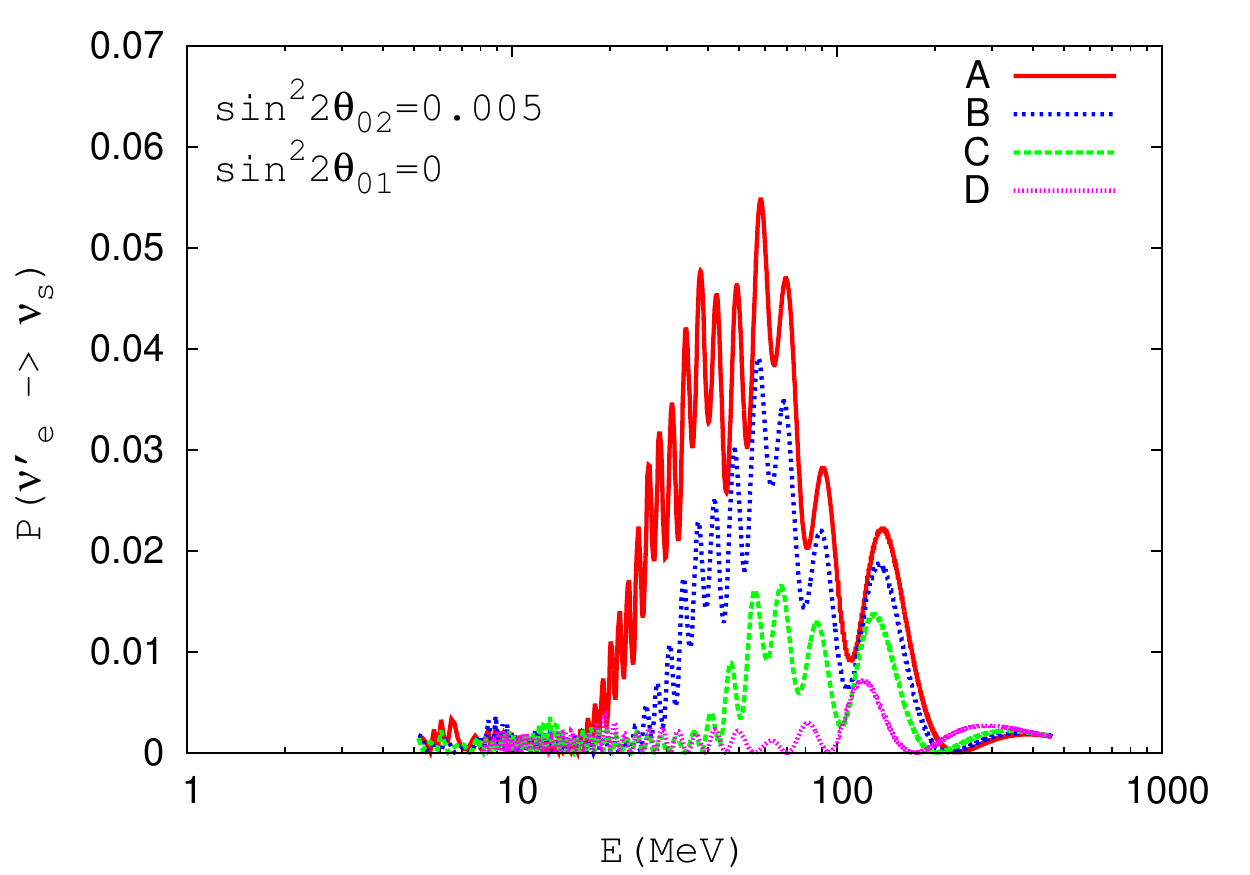}
\includegraphics[scale=1,width=8.2cm]{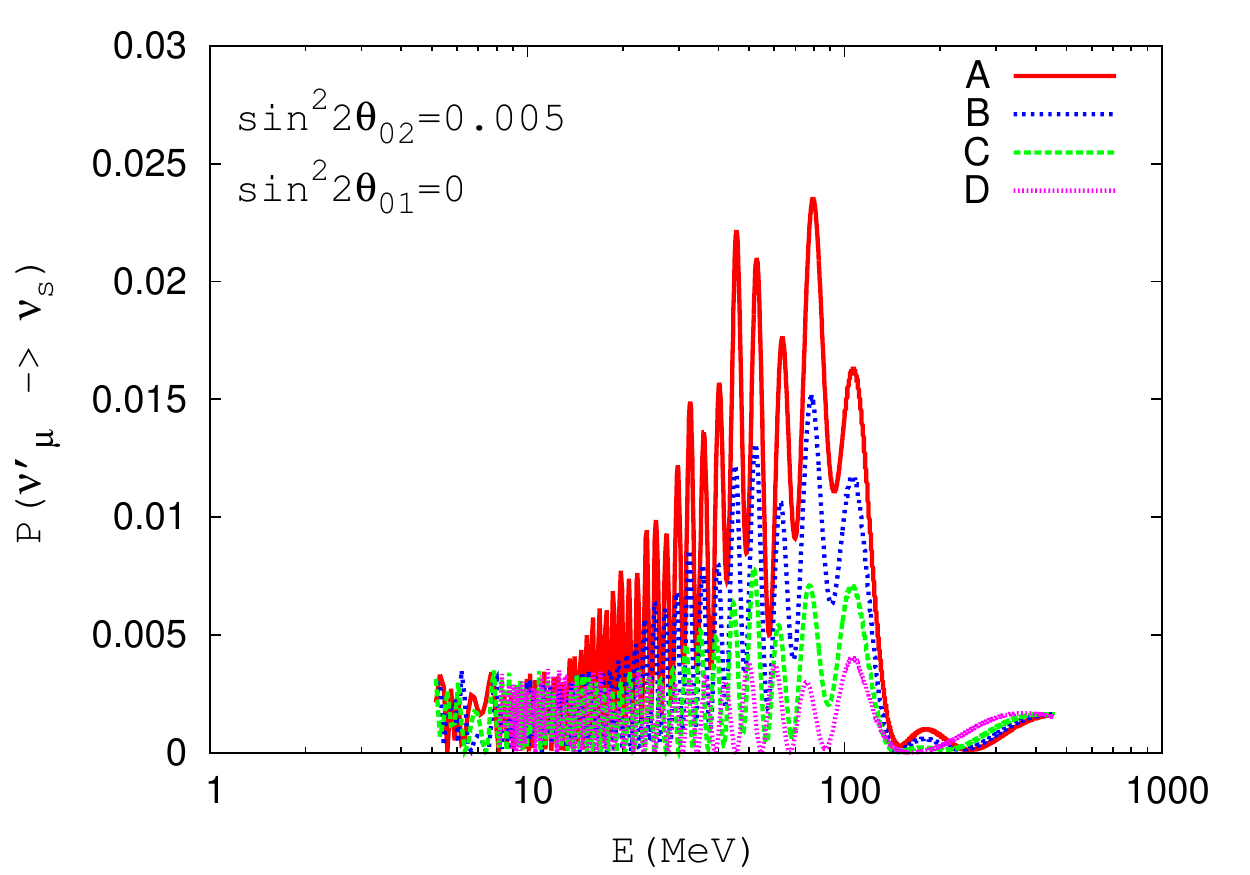}
\end{tabular}
\end{center}
\caption{$\nu'_e \to \nu_s$ and $\nu'_\mu\to \nu_s$ conversion
probability versus energy. Lines A, B, C and
D correspond to $\Delta {\widehat m}^2=0.5\times 10^{-5}$ eV$^2$,
$\Delta {\widehat m}^2=0.7\times 10^{-5}$ eV$^2$,
$\Delta {\widehat m}^2=1.0\times 10^{-5}$ eV$^2$,
$\Delta {\widehat m}^2=1.5\times 10^{-5}$ eV$^2$ separately.
$L=12000$km.}
\label{figurea}
\end{figure}

In Fig. \ref{figurea} we show the variation of
the probability of $\nu'_{e,\mu} \to \nu_s$ conversion
with respect to $\Delta {\widehat m}^2_{01}$. We can see that
as $\Delta {\widehat m}^2_{01} $ increases from the value
$0.5\times 10^{-5}$ eV$^2$, the amplitude of the
conversion probability decreases and for 
$\Delta {\widehat m}^2_{01} > 1.0\times 10^{-5}$eV$^2$
the probability is maximally around $1\%$. 
We see that the resonant conversion disappears for a relatively
large value of $\Delta {\widehat m}^2_{01}$.

For solar neutrinos one can similarly show that 
if $n_n/n_e \gsim 0.9$ in the Sun the level crossing would disappear.
For real matter density in the Sun which has a negligible neutron number
density, the level crossing and the MSW resonance indeed exist.

{\bf Formulation of oscillation of super-light sterile neutrino in the Earth}

In this section we present a perturbation theory describing the
oscillation of super-light sterile neutrino with active neutrinos
in the Earth matter. This formulation uses
the baseline averaged potential (\ref{poten})
and its basic strategy is the same as that presented for
oscillation among active neutrinos in ~\cite{Liao}.
We will show that the leading term in the theory, which is analytic,
gives a qualitatively good description of the
flavor conversion of the super-light sterile neutrino
$\nu_s$ with the active neutrinos. Including the first order
correction this perturbation theory is precise to describe the
flavor conversion between $\nu_s$ and $\nu'_{e,\mu}$.
This justifies the use of the average potential (\ref{poten})
in the discussion of the level crossing of energy levels
in the last section.
The theory is detailed below.

\begin{figure}[tb]
\begin{center}
\begin{tabular}{cc}
\includegraphics[scale=1,width=8cm]{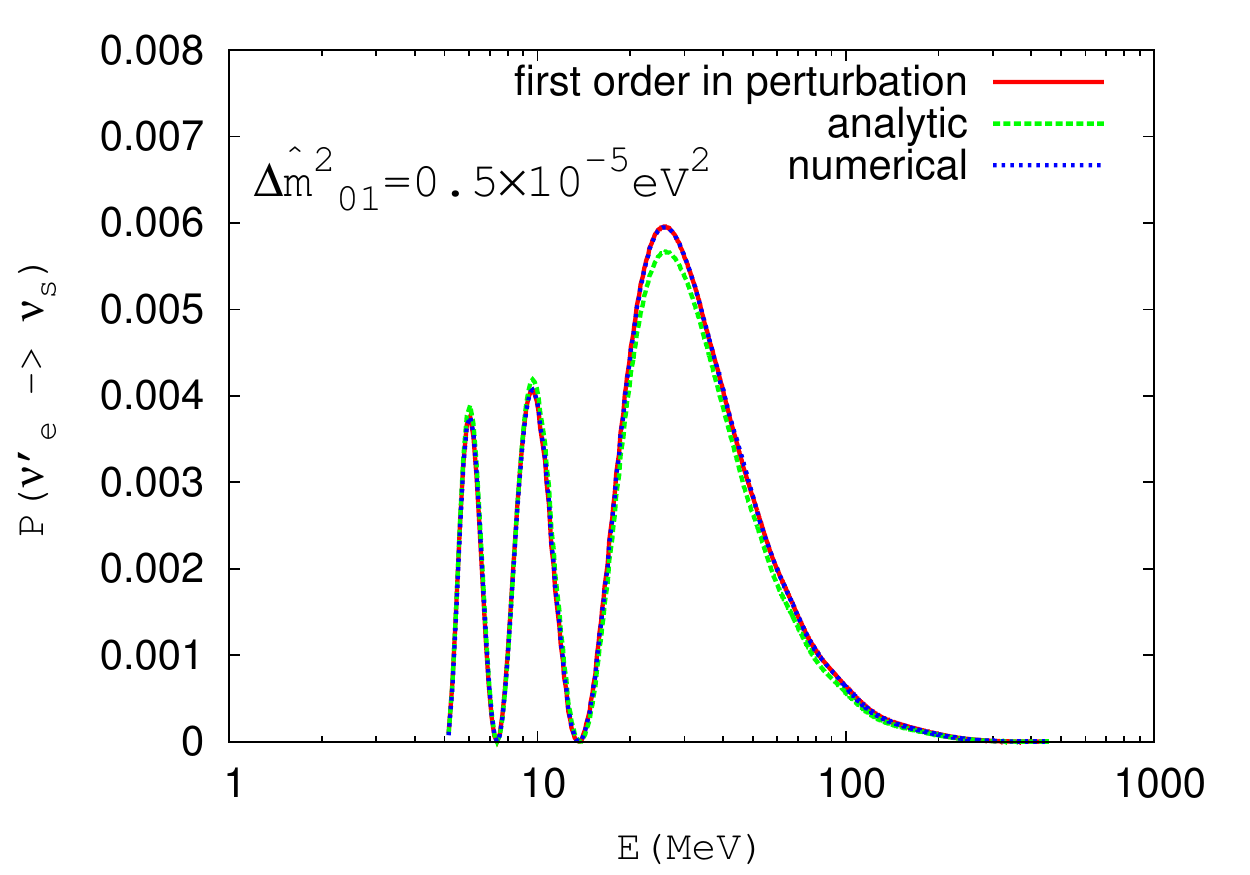}
\includegraphics[scale=1,width=8cm]{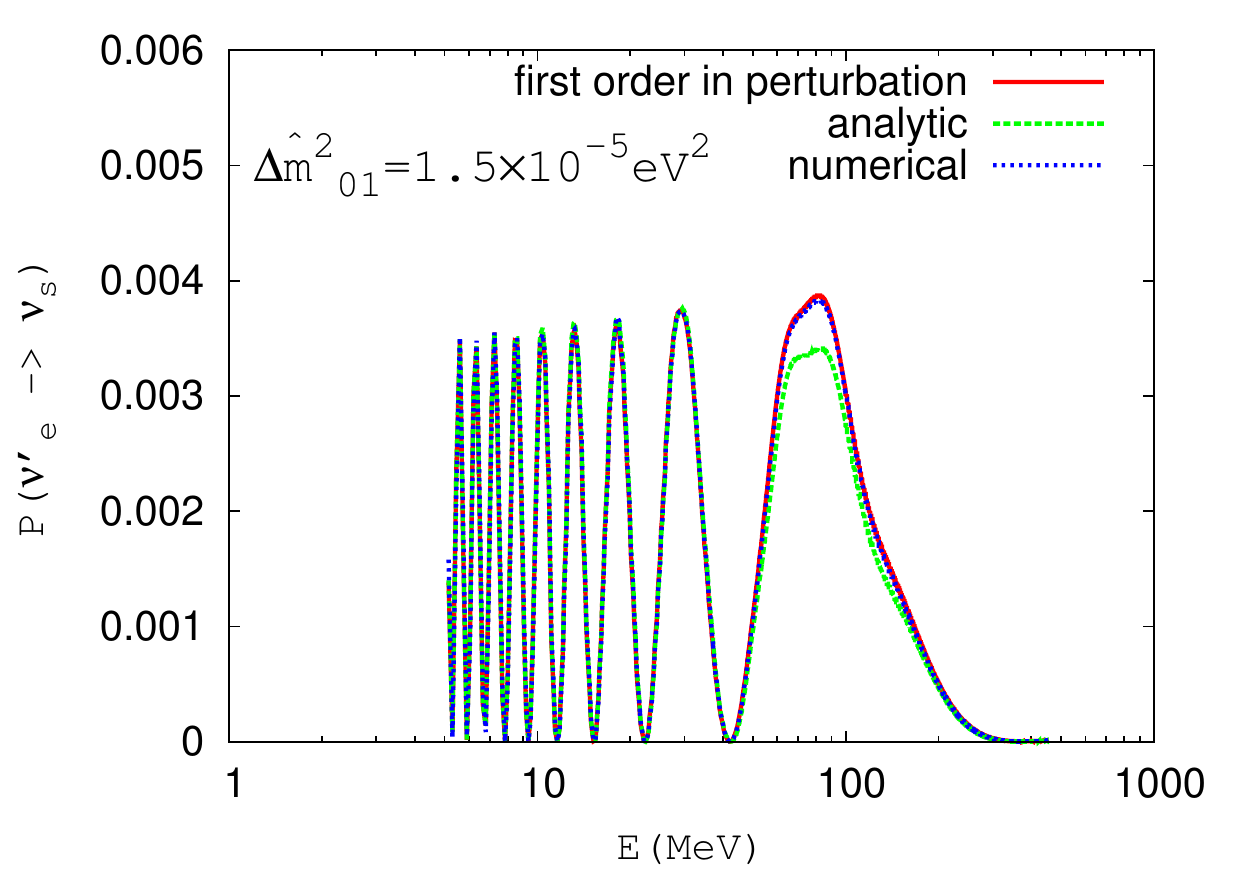}
\\
\includegraphics[scale=1,width=8cm]{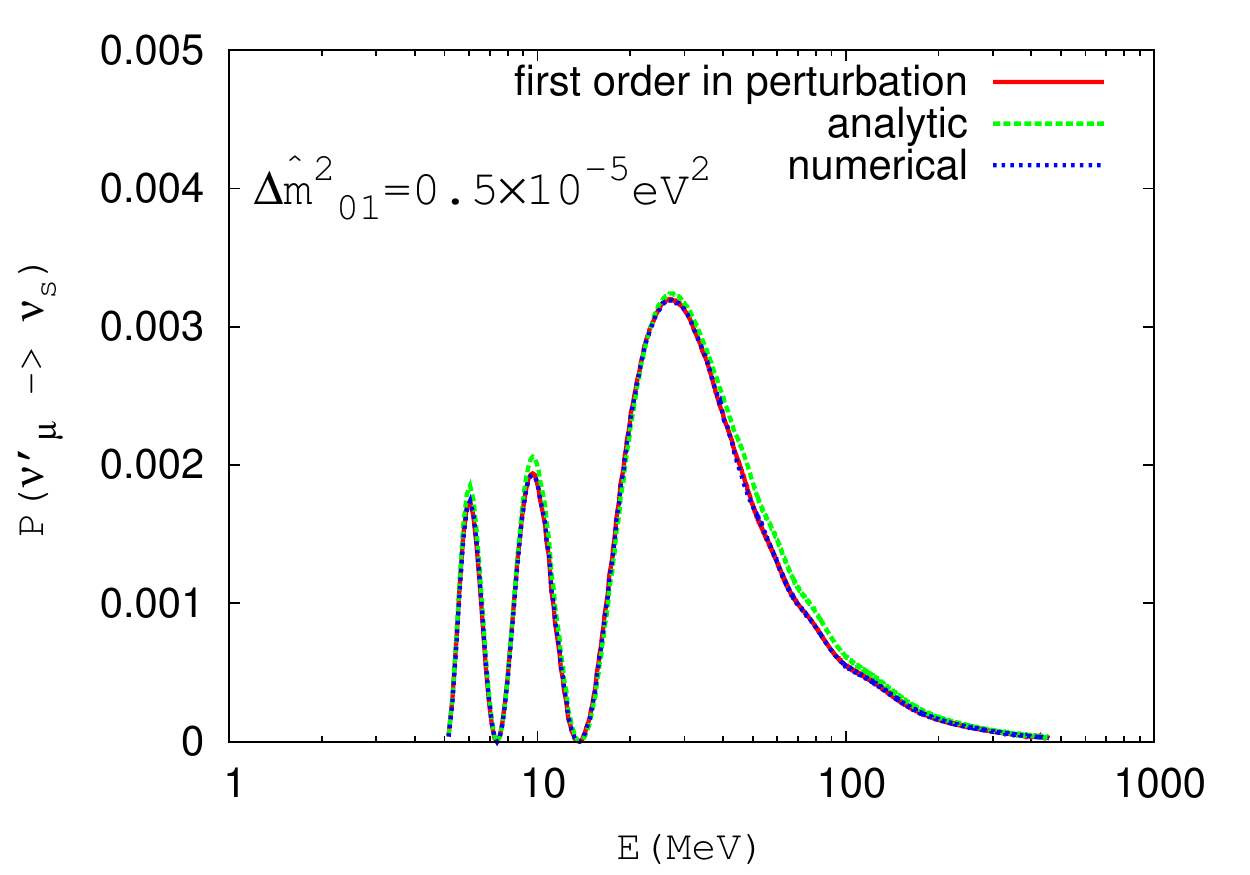}
\includegraphics[scale=1,width=8cm]{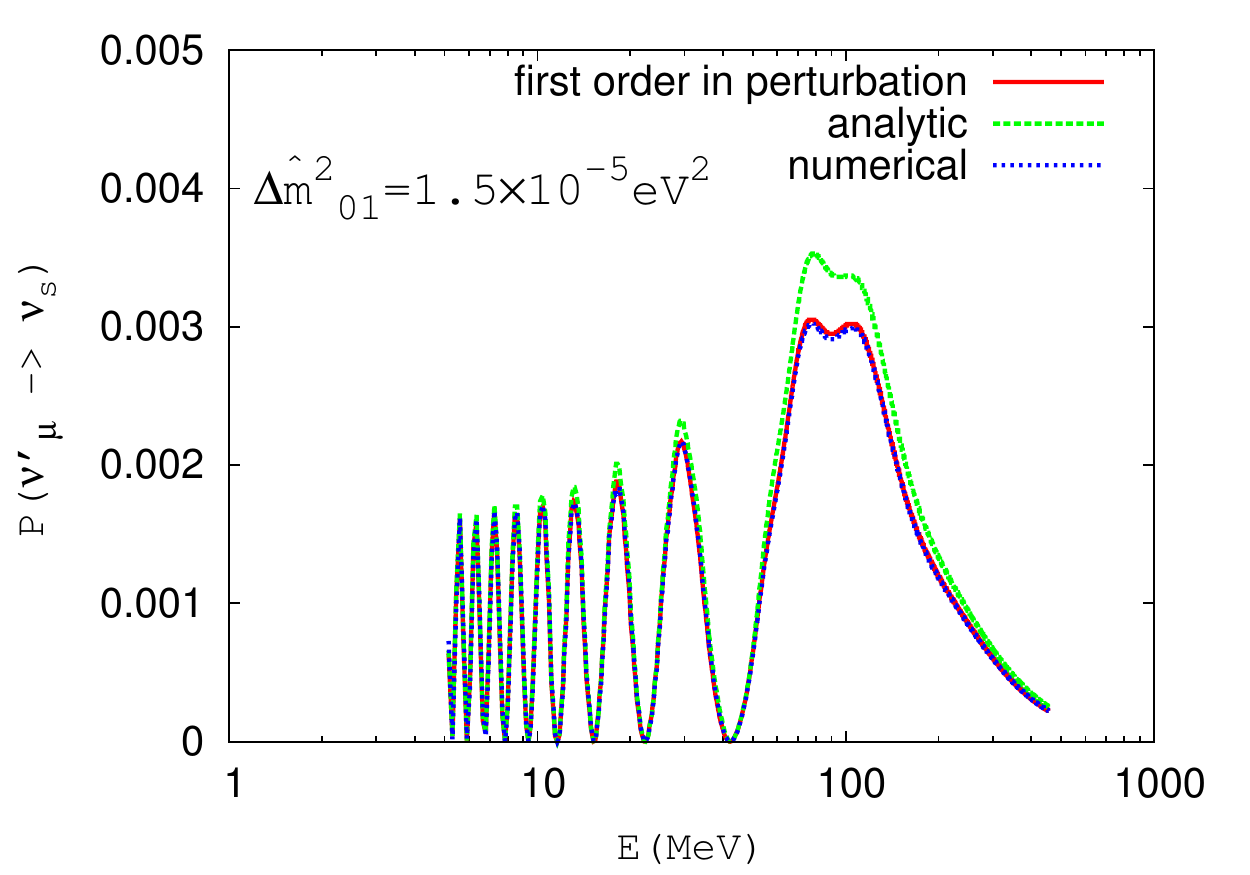}
\end{tabular}
\end{center}
\caption{$P(\nu'_e,\nu'_\mu \to \nu_s)$ versus energy for $L=8000$ km.
$\sin^2 2\theta_{01}=0.005$, $\theta_{02}=0$.}
\label{figure5}
\end{figure}

\begin{figure}[tb]
\begin{center}
\begin{tabular}{cc}
\includegraphics[scale=1,width=8cm]{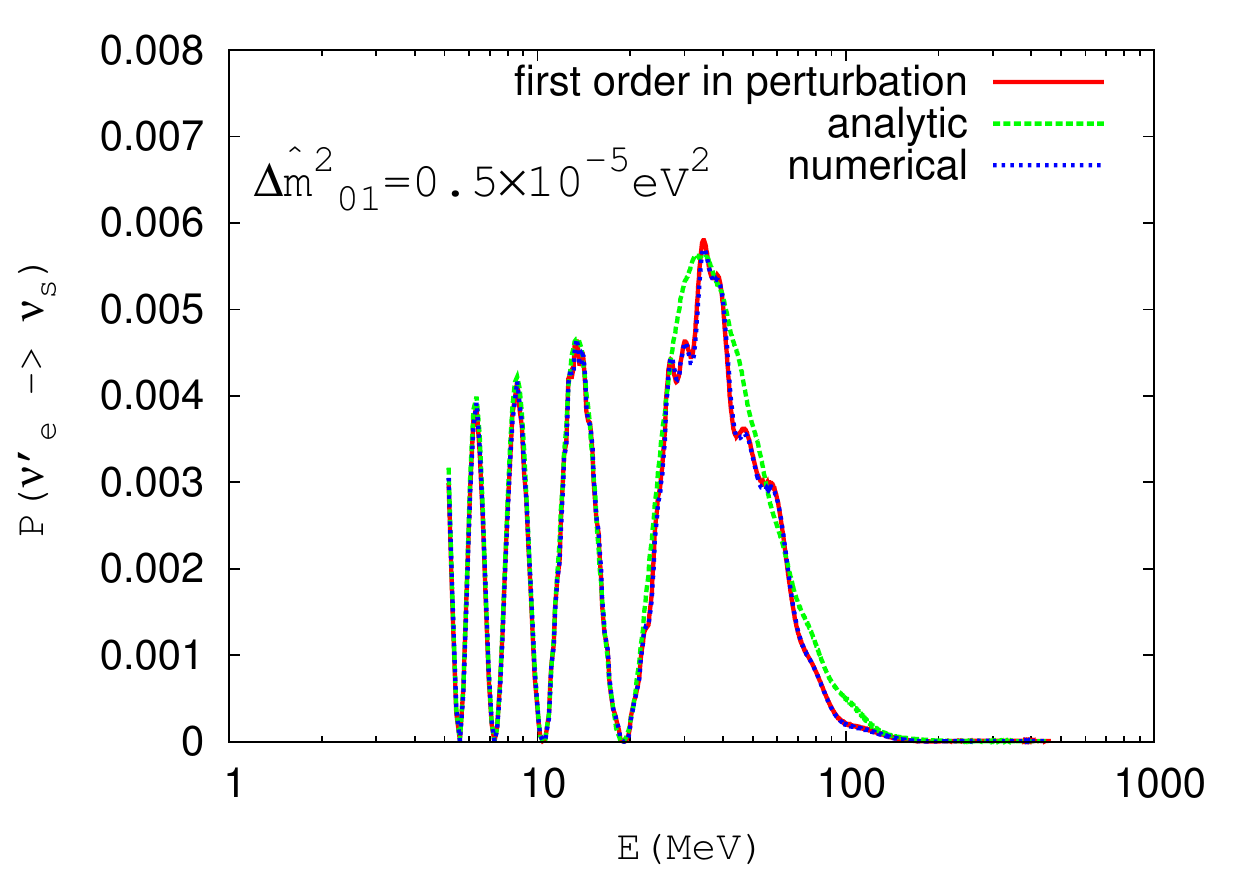}
\includegraphics[scale=1,width=8cm]{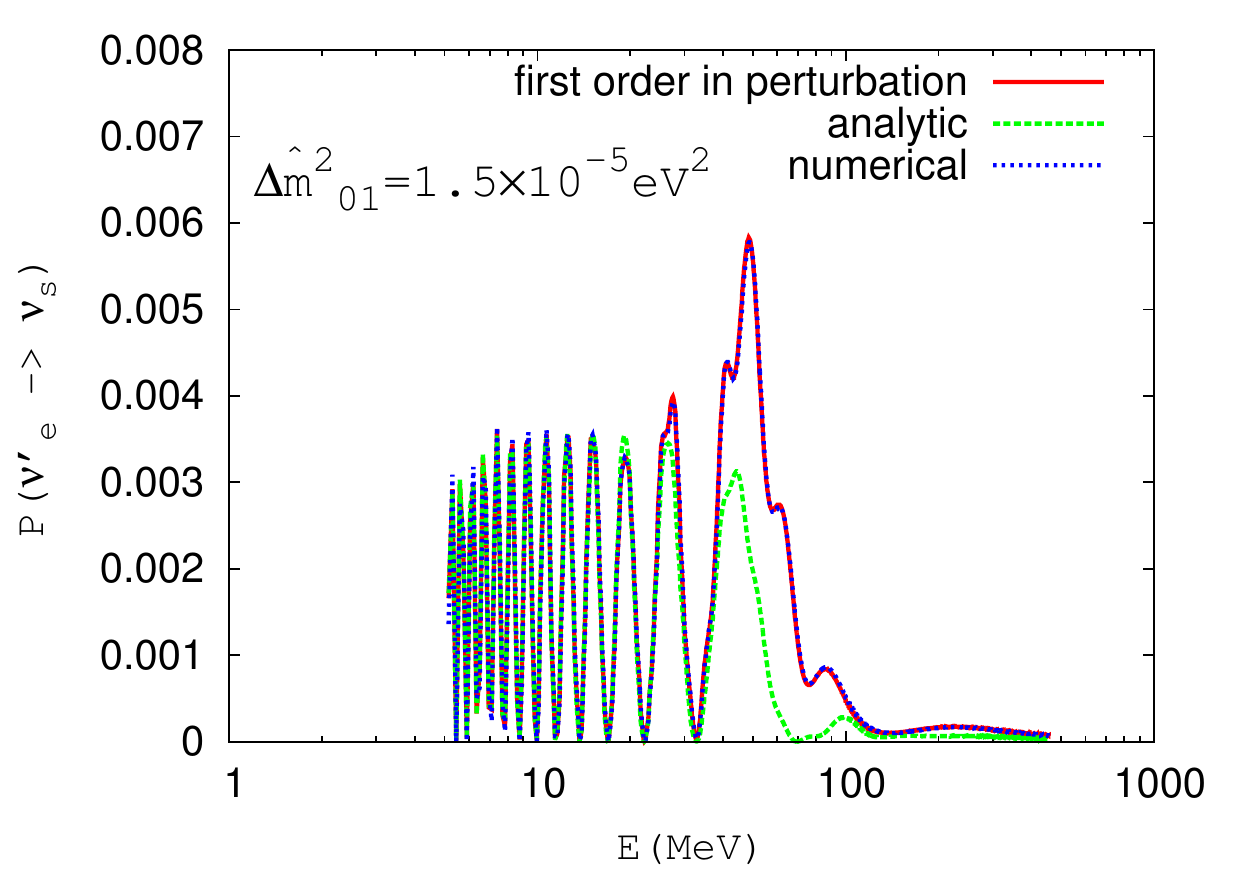}
\\
\includegraphics[scale=1,width=8cm]{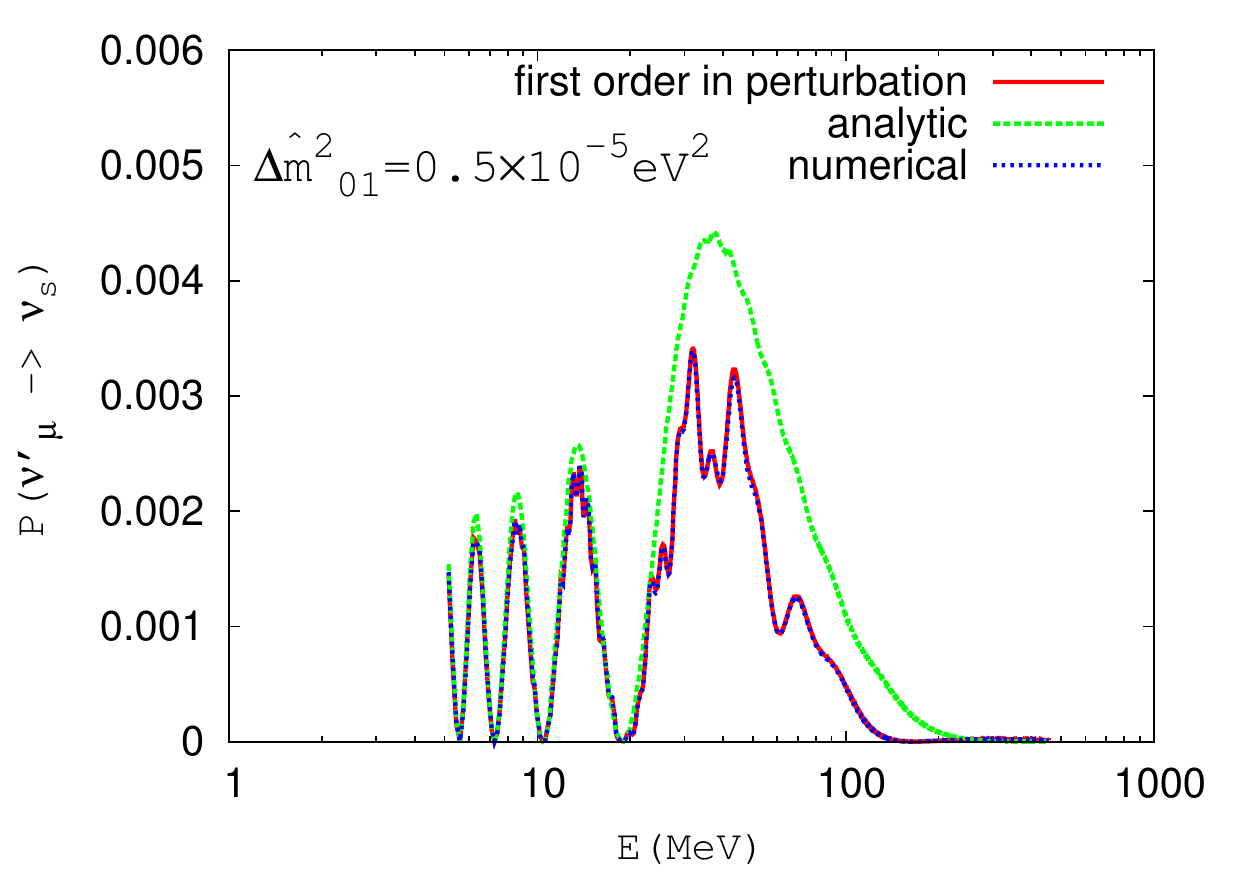}
\includegraphics[scale=1,width=8cm]{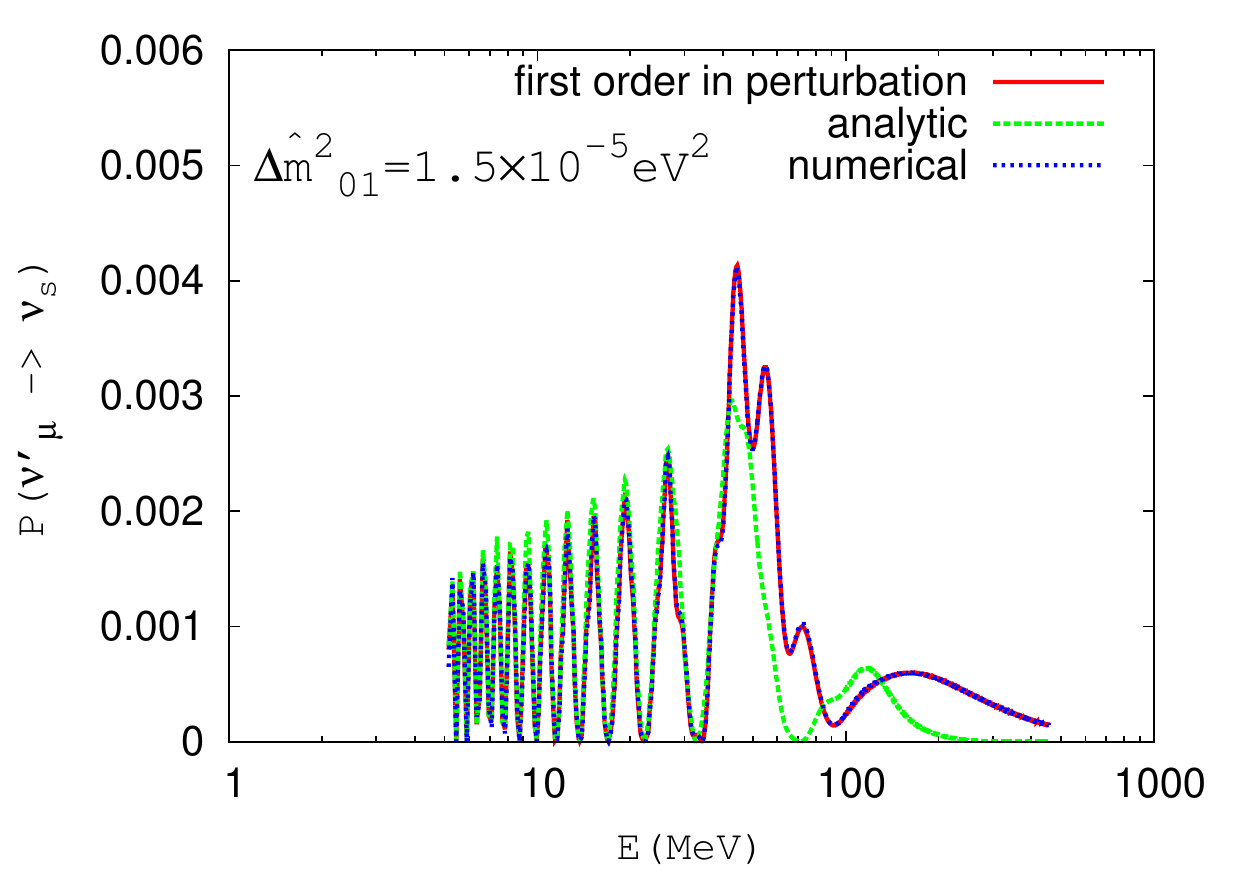}
\end{tabular}
\end{center}
\caption{$P(\nu'_e, \nu'_\mu \to \nu_s)$ versus energy for $L=12000$ km.
$\sin^2 2\theta_{01}=0.005$, $\theta_{02}=0$.}
\label{figure6}
\end{figure}

Similar to the formulation in ~\cite{Liao}, we can introduce
an average potential for a trajectory of neutrino with a baseline
length $L$ in the Earth
\bea
{\bar V}= \textrm{diag}\{0, {\bar V}_e+{\bar V}_n, {\bar V}_n \}
=\frac{1}{L} \int^L_0 ~dx~ {\widehat V}(x), \label{poten0}
\eea
where ${\bar V}_e$ has been introduced in (\ref{poten}) and
${\bar V}_n$ is similarly defined. The Hamiltonian (\ref{H3-0})
can be rewritten using (\ref{poten0}) as
\bea
{\widehat H}(x)={\bar H}+ \delta {\widehat H}(x),
\eea
where
\bea
{\bar H}={\hat U} {\widehat H}_0 {\hat U}^\dagger+{\bar V},
\eea
and $\delta {\widehat H}={\widehat H}-{\bar H}$ which equals to $\delta {\widehat V}$
\bea
\delta {\widehat H}
=\delta {\widehat V}={\widehat V}-{\bar V}.
\eea

Introducing a mixing matrix $U_m$ in matter which
diagonalizes ${\bar H}$:
 \bea
 {\bar H} U_m = U_m \frac{1}{2 E} \Delta, ~~
 \Delta=\textrm{diag}\{\Delta^1, \Delta^2, \Delta^3\},
 \label{defU}
 \eea
where $\frac{1}{2 E} \Delta^i(i=1,2,3)$ are three eigenvalues of ${\bar H}$,
we are ready to solve the evolution problem in (\ref{H3}) perturbatively
in an expansion in $\delta {\widehat V}$.
We first solve the evolution governed by ${\bar H}$ and obtain
the contribution of $\delta V$ using perturbation in $\delta V$.
Keeping result of first order in $\delta V$ we obtain
 \bea
 {\widehat \nu}'(L)&&=M(L) {\widehat \nu}'(0), \label{evol2} \\
 M(L) &&=U_m ~e^{- i \frac{\Delta}{2 E} L}(1- i C) ~U^\dagger_m,
 \label{evol2b}
  \eea
 where $C$ is a $3\times 3$ matrix accounting for the non-adiabatic
 correction
 \bea
 C=\int^L_0 dx ~e^{i \frac{\Delta}{2 E} x} U^\dagger_m \delta V(x) U_m
  e^{-i \frac{\Delta(x)}{2 E} x} \label{evol3}.
 \eea
It is clear that $C^\dagger =C$ holds. One can see that
 \bea
 C_{jj} &&=\int^L_0 dx ~(U^\dagger_m \delta V(x) U_m)_{jj}=0, \label{C1} \\
 C_{jk} &&=\int^L_0 dx ~e^{i \frac{\Delta^j-\Delta^k}{2 E} x} (U^\dagger_m
 \delta V(x) U_m)_{jk}, ~~j\neq k . \label{C2}
 \eea
 Eq. (\ref{C1}) is guaranteed by Eq. (\ref{poten0}).
 $|C_{jk}| \ll 1 (j\neq k)$ should be satisfied if this is a good perturbation
 theory. One of the virtues of this perturbation theory is that Eq. (\ref{C1})
  guarantees that the oscillation phase is correctly reproduced.

\begin{figure}[tb]
\begin{center}
\begin{tabular}{cc}
\includegraphics[scale=1,width=8cm]{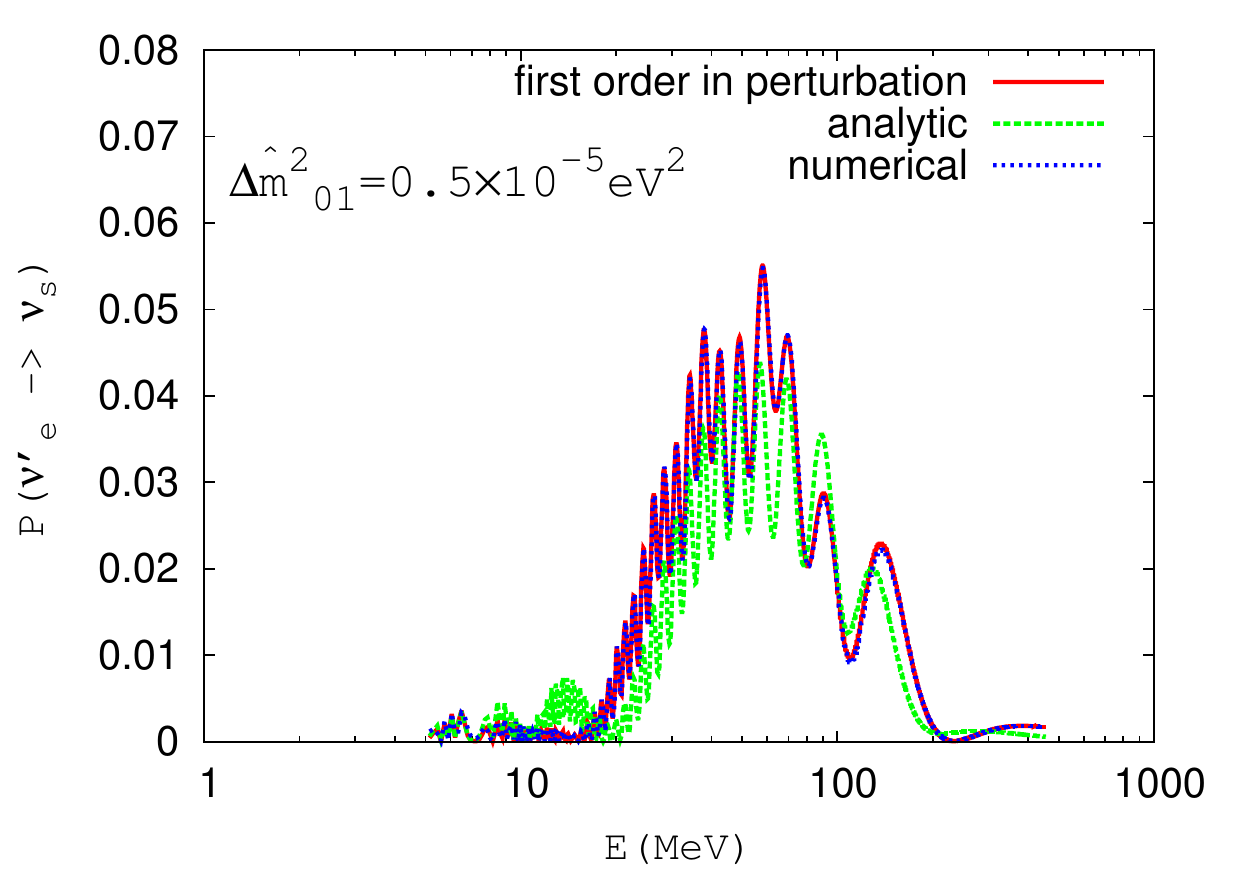}
\includegraphics[scale=1,width=8cm]{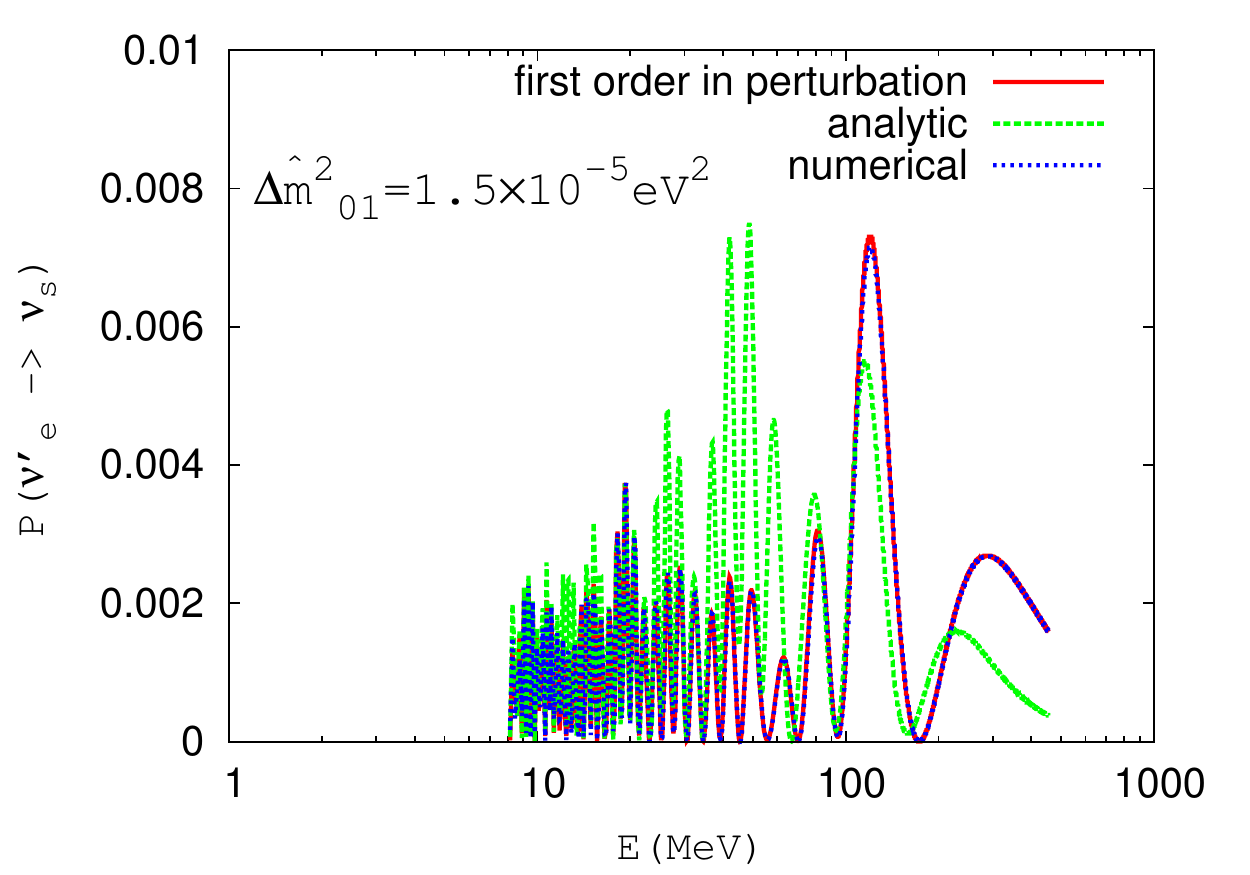}
\\
\includegraphics[scale=1,width=8cm]{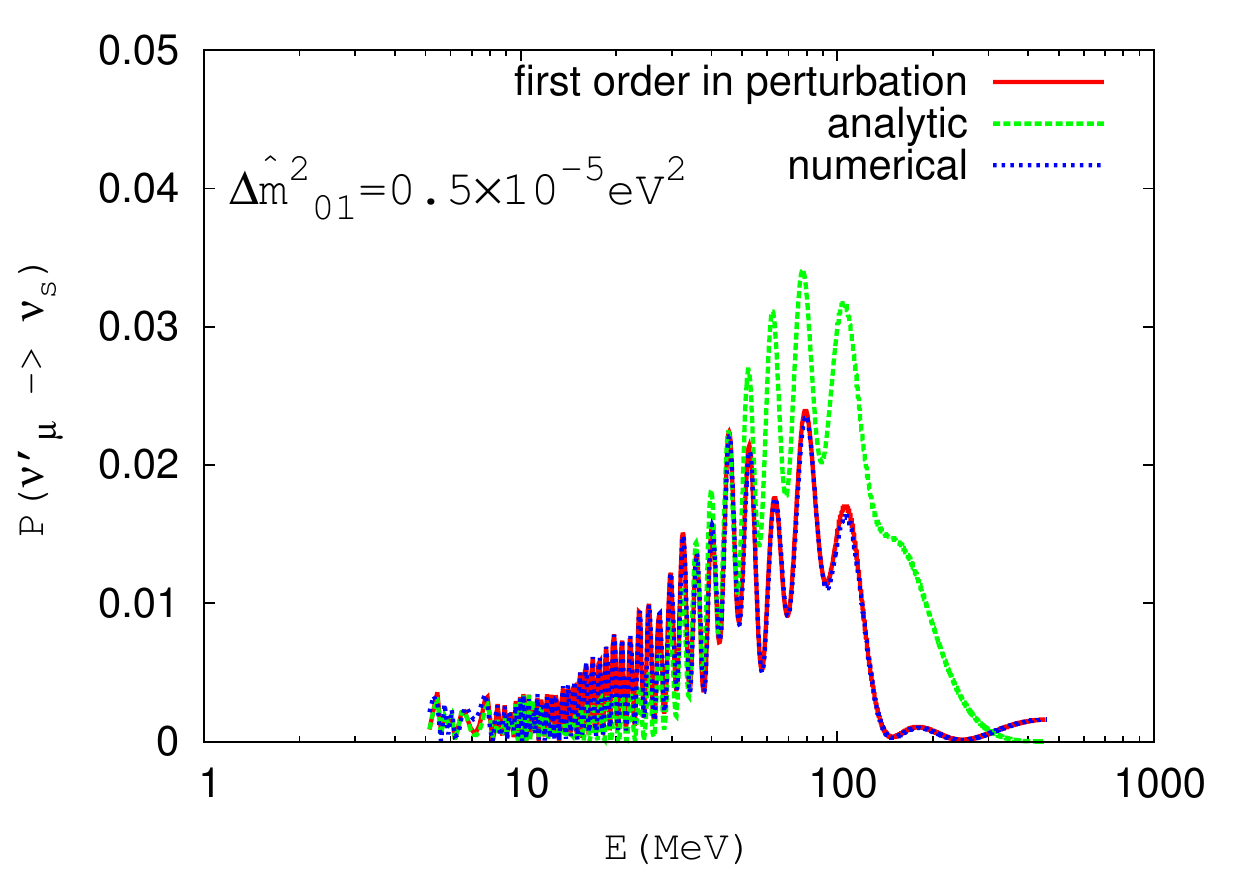}
\includegraphics[scale=1,width=8cm]{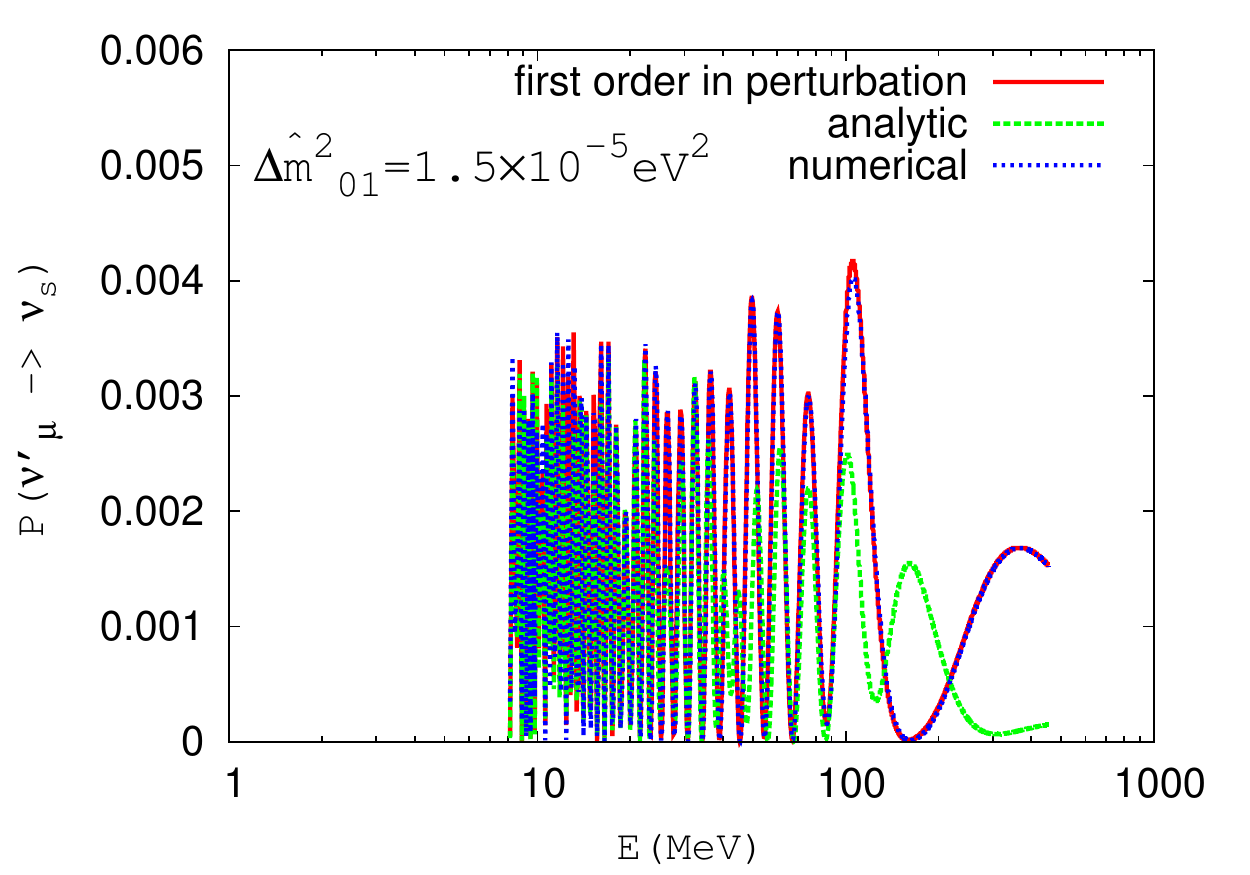}
\end{tabular}
\end{center}
\caption{$P(\nu'_e,\nu'_\mu \to \nu_s)$ versus energy for $L=12000$ km.
$\sin^2 2\theta_{02}=0.005$, $\theta_{01}=0$.}
\label{figure7}
\end{figure}

\begin{figure}[tb]
\begin{center}
\begin{tabular}{cc}
\includegraphics[scale=1,width=8cm]{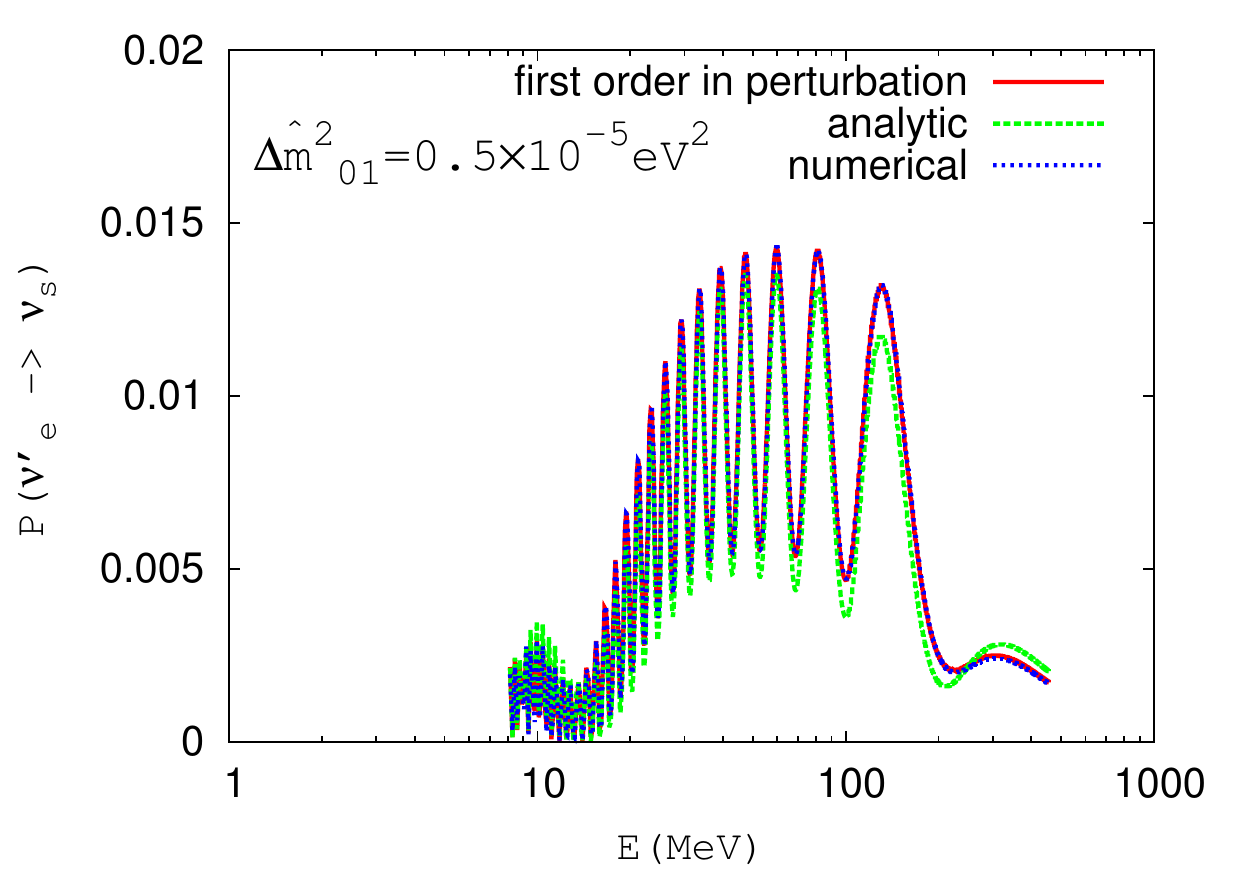}
\includegraphics[scale=1,width=8cm]{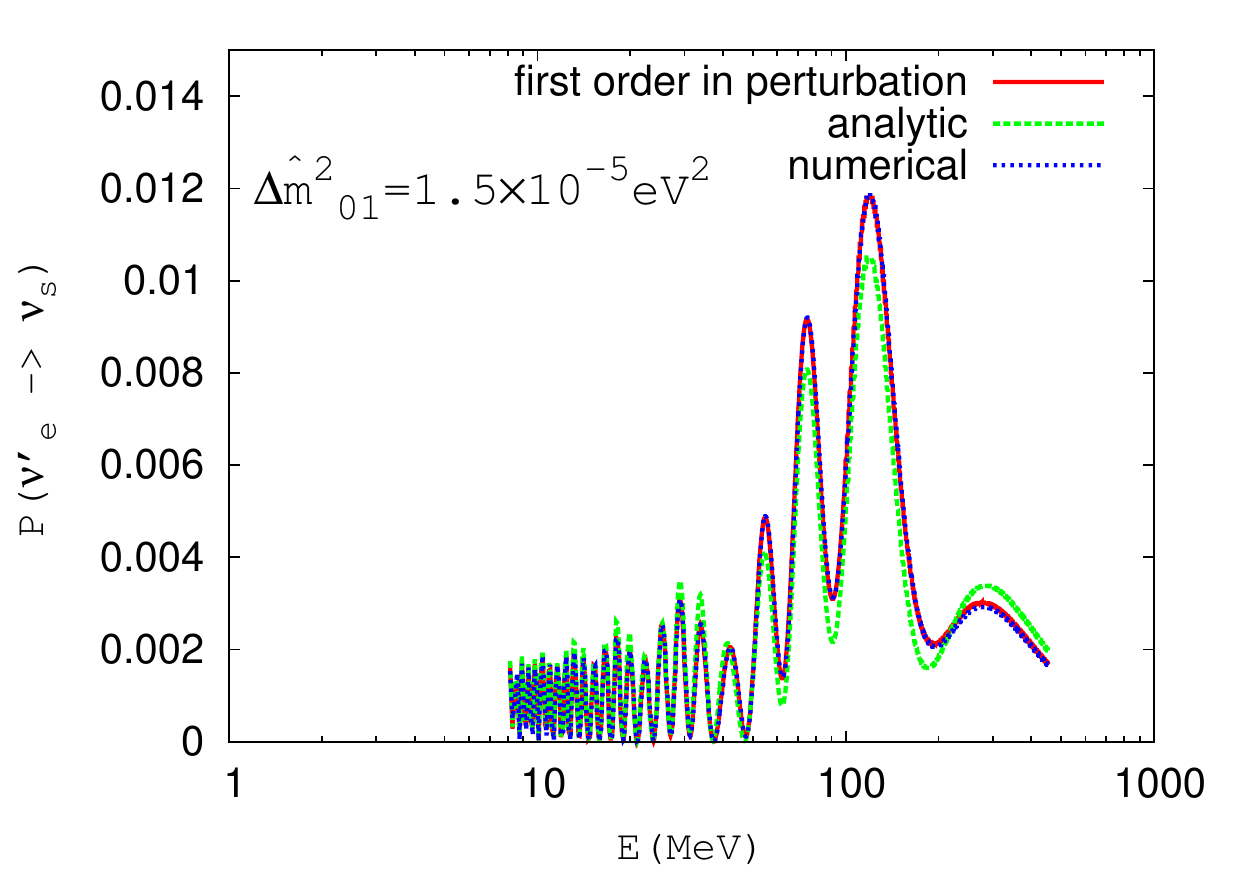}
\\
\includegraphics[scale=1,width=8cm]{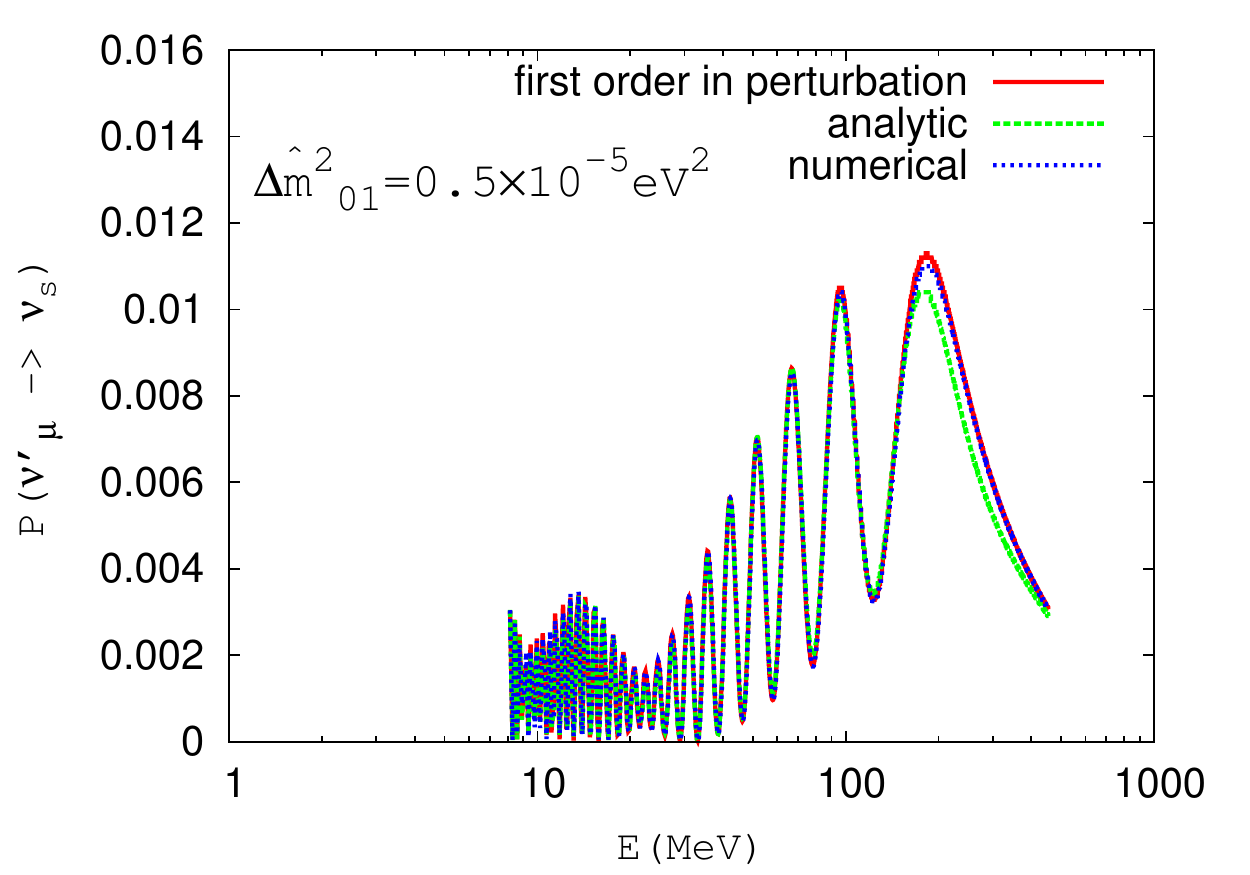}
\includegraphics[scale=1,width=8cm]{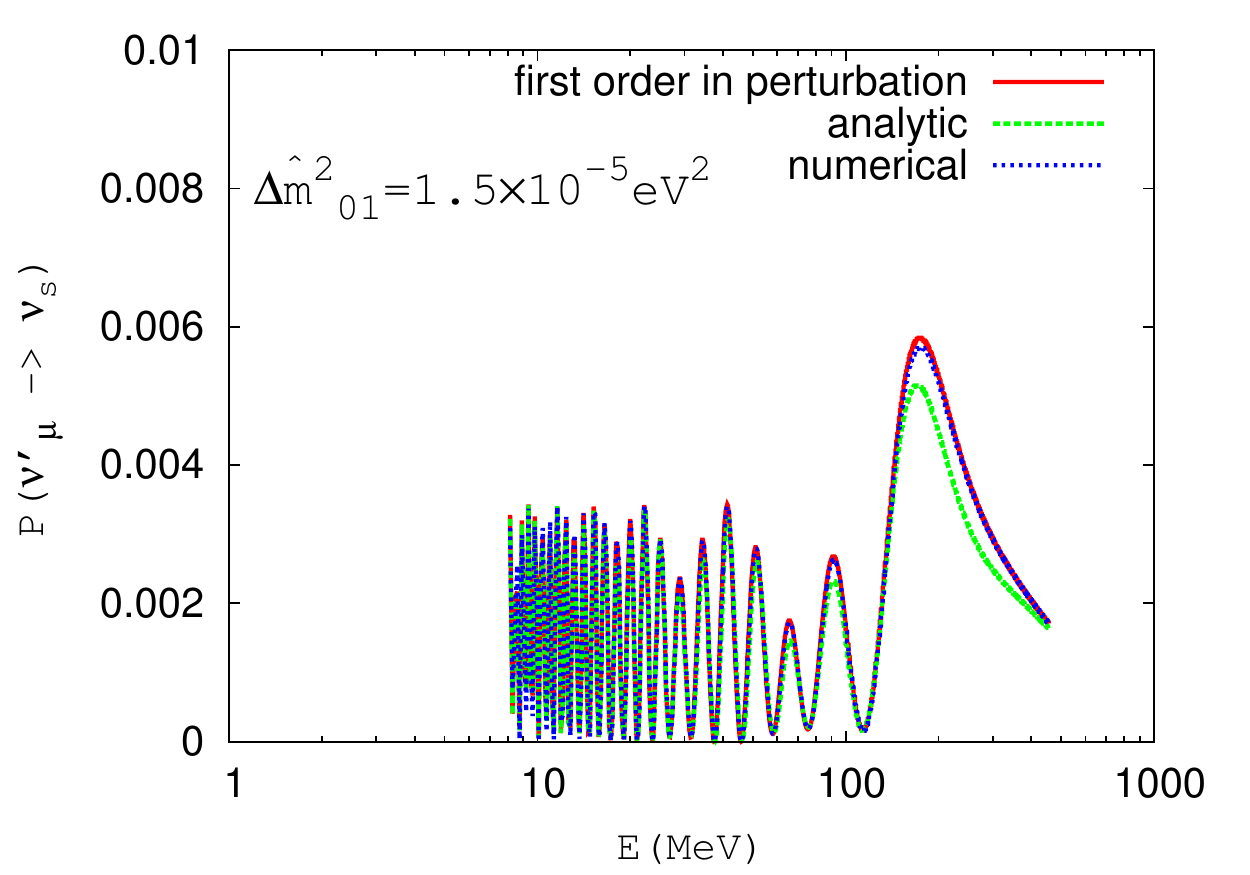}
\end{tabular}
\end{center}
\caption{$P(\nu'_e,\nu'_\mu \to \nu_s)$ versus energy for $L=8000$ km.
$\sin^2 2\theta_{02}=0.005$, $\theta_{01}=0$.}
\label{figure8}
\end{figure}

In Fig. \ref{figure5} we give plots for $\nu_s-\nu'_e$ conversion
versus energy for $L=8000$ km. We give zero-th order
results calculated using analytic formula, i.e. using (\ref{evol2}) 
with $C$ switched off. We also present the result including
the first order correction (\ref{evol3}). The line for
numerical result is calculated using the PREM Earth density profile~\cite{PREM}.
We can see that the zero-th order result using
trajectory averaged potential reproduces
the oscillation phase very well, but not the magnitude. Including
the first order correction calculated using (\ref{evol3}), 
this formalism reproduces very well the
oscillation phase and magnitude of the flavor conversion.

When $L > 10690$ km neutrinos cross the core of the Earth. Large density 
jump between the core and the mantle makes the above simple version of
the perturbation theory not as precise as shown in Fig. \ref{figure5}. 
We can improve the approximation by dividing the whole trajectory
into three parts with parts 1 and 3 in the mantle and part 2 in the core
of the Earth. The evolution matrix can be written as
\bea
M = M_3 M_2 M_1,  \label{evol4}
\eea 
where $M_2$ is the evolution matrix in the core and $M_{1,3}$ are evolution 
matrices in the mantle. $M_i(i=1,2,3)$ are calculated similar to
(\ref{evol2}) but with average potentials evaluated in the corresponding
part of the trajectory in the mantle and in the core separately~\cite{Liao}.
In Fig. \ref{figure6} we give plots for
an example of this case. We can see that the result calculated using
the improved perturbation theory (\ref{evol4}) agrees well with the numerical
result. The oscillation phase and the magnitude of the flavor conversion
are very well reproduced in the perturbation theory. We also give
the analytic result calculated using the potential averaged over the
whole trajectory (\ref{poten0}). We can see that this analytic result
correctly reproduces the oscillation phase of the $\nu_s-\nu'_e$ conversion.

In Fig. \ref{figure7} and \ref{figure8} we give plots of
$\nu'_{e,\mu} \to \nu_s$ oscillation similar to
Fig. \ref{figure5} and \ref{figure6}, but for non-zero $\theta_{02}$.
Similarly, we can see that the analytic result gives a good account
of the phase of the oscillation of the super-light sterile neutrino
with active neutrinos. The peterubation theory gives not only
reproduces the oscillation phase but also the magnitude of the
flavor conversion very well.

We note that the above discussion does not mean that
the perturbation theory presented in this article give a precise
description for $1-2$ oscillation. The theory presented
in this article indeed give a qualitatively good description
for $1-2$ oscillation for energy $\gsim 20$ MeV, but not precisely.
Actually for an energy around 10 MeV the structure of the Earth matter
shows up in $1-2$ oscillation~\cite{Liao0} and 
the perturbation theory using average potential is not a
precise description.
This perturbation theory gives a precise description of $1-2$ oscillation
for energy $\gsim 500$ MeV~\cite{Liao}.

We can see from the above discussion that the zero-th order result
calculated using
the trajectory averaged potential, Eq. (\ref{poten0}), always
give a correct account of the oscillation phase of the $\nu_s-\nu'_{e,\mu}$
conversion. Although the zero-th order result does not give
a precise description of the magnitude of conversion, it encodes
the major properties of the flavor conversion in the Earth. This justifies
the use of the average potential in the discussion of the 
level crossing and the
disappearance of the resonance in the last section.

{\bf Conclusion:}

In summary we have made a detailed study of the flavor
conversion of the super-light sterile neutrino with
active neutrinos in Earth matter. A super-light sterile neutrino,
with a mass squared difference $\Delta m^2_{01}\approx (0.5-2)\times 10^{-5}$ eV$^2$
and a small mixing angle $\theta_{01}$ or $\theta_{02}$, can 
oscillate with electron neutrinos in the Sun through a MSW resonance
and can help to explain
the absence of the upturn of the solar boron neutrino spectrum
at energy $\lsim 4$ MeV. One would naively expect that
a similar resonant conversion should also happen when
neutrinos pass through the Earth. In this article we have shown that for
$\Delta m^2_{01} \gsim 1\times 10^{-5}$ eV$^2$
this naively expected resonant conversion disappears completely 
and for smaller value of $\Delta m^2_{01}$ there is still an enhancement of the flavor
conversion but the conversion probability is at most a few percent.

We have shown that the absence or the suppression
of the resonant conversion is because of the presence of the potential $V_n$
which arises from neutral current interaction of active 
neutrinos with neutrons in matter. The neutron number density in the Sun
is negligible comparing to the electron density and the
effect of $V_n$ is basically switched off. On the other hand, the
neutron number density in the Earth roughly equals to the electron
number density and the effect of $V_n$ can play
important role. We have shown that the naively expected level crossing
of energy levels of neutrinos disappears when including the
effect of $V_n$ in Earth matter. In particular, we find that
for $\Delta m^2_{01}=(0.5-2)\times 10^{-5}$ eV$^2$
the energy of the super-light neutrino is always in-between the
energies of $\nu_1$ and $\nu_2$ neutrinos and there is
no crossing of energy levels among these neutrinos. For $\Delta m^2_{01}$
around $0.5\times 10^{-5}$ eV$^2$ the energies of the super-light
sterile neutrino and the active neutrino $\nu_1$ have a chance
to be close to each other when the energy is
around 60 MeV and hence create a resonant conversion. However this
resonant enhancement of the flavor conversion is significantly 
suppressed comparing to the case when $V_n$ is switched off
and the flavor conversion probability is at most a few percent.
Furthermore, the position of the resonance is shifted to an energy
around $60$ MeV which makes this scenario easily be able to
escape possible contraints
coming from measurement of Earth matter effect
in solar neutrino experiments.

Apparently the scenario of super-light sterile neutrino is difficult to 
test in ground-based experiment of neutrino oscillation. We have shown
that the conversion of $\nu'_{e,\mu}\to\nu_s$ is maximally a few percent
and the maximal conversion happens for energy around $60$ MeV. This
energy range is well beyond that of the solar and supernovae neutrino
spectrum. The only possibility is to test it in very low energy
atmospheric neutrino data. However, this is also pretty difficult because
the conversion probability is maximally a few percent but there are
quite a lot of uncertainties in low energy atmospheric neutrinos.

For completeness of our discussion we have also shown a perturbation theory
which makes use of an baseline averaged potential in developing the theory.
This theory can very well describe the flavor conversion of the super-light
sterile neutrino with active neutrinos in Earth matter. This justifies the
use of the baseline averaged potential in the discussion of the
level crossing of neutrinos. We have also
discussed the reduction of the flavor conversion in the $4\nu$ framework
to a $3\nu$ framework.

The outcome of the present research is interesting. A MSW resonance
is needed in explaining the absence of the upturn of the solar boron
neutrino at energy $\lsim 4$ MeV~\cite{deHS}. However,
a resonant conversion of this super-light sterile neutrino
with active neutrino is dangerous when confronting this scenario with the
experiments which can test the Earth matter effect, e.g. the solar and
atmospheric neutrino experiments. Interestingly, we find that
the effect of $V_n$ in the Earth makes the resonant conversion
disappeared or significantly suppressed and the scenario of the
super-light sterile neutrino can pass through test of Earth matter effect
in neutrino oscillation experiments.
On the other hand, 
the effect of $V_n$ in the Sun is actually switched off because of the negligible
neutron number density and 
this enables the MSW resonance in the Sun.
This is exactly the situation welcomed when confronting the scenario
of the super-light sterile neutrino with oscillation experiments. 
It's interesting to see that it is the different situation in
the Sun and in the Earth that makes $V_n$ effectively turned off and turned
on respectively. This observation 
seems to suggest that the super-light sterile neutrino is a natural choice
made by the Sun and the Earth rather than by the authors of Refs. \cite{deHS0,deHS}.
It makes the scenario of the super-light sterile neutrino very interesting.

\acknowledgments
This work is supported by National Science Foundation of
 China(NSFC), grant No.11135009, No. 11375065 and Shanghai Key Laboratory
 of Particle Physics and Cosmology, grant No. 11DZ2230700.


\begin{thebibliography}{99}
\bibitem{deHS1}
For lower bound on $\Delta m^2_{01}$ we take $0.5\times 10^{-5}$ eV$^2$
which comes from a discussion of Fig. 7 in Ref. \cite{deHS}. Taking the
more conservative lower bound $0.7\times 10^{-5}$ eV$^2$ does not change the
conclusion in this article.
\end{thebibliography}
\end{document}